\documentclass[prd,reprint,superscriptaddress,preprintnumbers,nofootinbib,amsmath,amssymb,aps]{revtex4-2}
\usepackage{booktabs}
\usepackage{graphicx}
\usepackage{multirow}
\usepackage{dcolumn}
\usepackage{color}
\usepackage[colorlinks=true,citecolor=blue,urlcolor=blue]{hyperref}
\usepackage{import}
\graphicspath{{./fig/}}
\begin{document}
\title{Constraints on the interacting holographic dark energy models: implications from background and perturbations data\\
}
\author{N.Nazari Pooya}
\thanks{Email: nazarip@basu.ac.ir}
\affiliation{Department of Physics, Faculty of Science, Bu-Ali Sina University, Hamedan 65178 016016, Iran}
\begin{abstract}
In this study, we employ a two-step method to analyze models of holographic dark energy (HDE) and interacting holographic dark energy (IHDE), incorporating three distinct dark energy (DE)-dark matter (DM) interaction terms. First, using the latest background dataset, we conduct a Markov chain Monte Carlo (MCMC) analysis to constrain the free parameters of the models. 
Then, we assess the models against each other using the key background parameters and compare them to the $\Lambda$CDM standard model. Our results show that at high redshifts, the equation of state (EoS) parameter related to the models for both homogeneous and clustered DE cases falls within the quintessence region. However, as we approach the present time, all models except HDE transition into the phantom region, and two models cross the phantom line earlier than others. In the next step, we focus on the evolution of perturbations in DE and DM. Using background and growth rate data, we constrain parameters including $\sigma_8$. We then investigate the evolution of the growth rate of matter perturbations, $f\sigma_8(z)$, and its deviation, $\Delta f \sigma_8$, from the $\Lambda$CDM model. The HDE model shows the best agreement with observational data, while other models predict varying growth rates compared to $\Lambda$CDM.
Finally, we demonstrate through Akaike and Bayesian information criteria (AIC and BIC) analysis that the compatibility of models with observational data depends on the type of data used, the DE-DM interaction term, and the assumptions regarding DE homogeneity and clustering. Our results suggest that homogeneous DE models yield more agreement with observational data than clustered DE models.
\end{abstract}
\maketitle
\section{Introduction}
Evidence gathered from various sources strongly supports the idea that the Universe's expansion is accelerating. This compelling evidence spans a wide array of measurements, including observations of supernovae type Ia (SnIa) \citep{Riess1998,Perlmutter1999,Kowalski2008}, cosmic microwave background (CMB) \citep{Jarosik2011, Komatsu2011, Planck Collaboration XIV2016, Aghanim et al2020}, baryon acoustic oscillations (BAO) \citep{Reid2012,Blake2011b,Percival2010,Tegmark2004,Cole2005}, distant high-redshift galaxy clusters \citep{WangSteinhardt1998a,Allen2004}, weak lensing surveys \citep{Fu:2008,Amendola:2007rr,Benjamin:2007}, and various other sources. Collectively, these diverse observations consistently reinforce the concept of an accelerating expansion, shedding light on the Universe's evolution and the crucial role of dark energy (DE).
Despite the robust support for the $ \Lambda$ cold dark matter ($ \Lambda$CDM)model provided by these observations, several challenges remain. The nature of DE itself continues to be shrouded in mystery, with its origin and properties still not fully elucidated. Issues such as the fine-tuning and the cosmic coincidence problems \citep{Winberg1989, SahniStarobinsky2000, Carroll2001, Padmanabhan 2003, Copeland et al. 2006, Peebles2003, Woodard2007, Durrer2008} raise fundamental questions about why DE dominates the Universe's energy density at the present epoch. Moreover, tensions surrounding parameters like $S_{8}$ \citep{Joudaki2018,Abbott2018,Basilakos2017} and $ H_{0}$ \citep{Abdalla2022,Kazantzidis2019,Valentino2021,Schoneberg2022,Alestas2020,Shah2021}   further complicate the landscape, presenting both theoretical and observational challenges.

When we observe the formation of cosmic structures on smaller scales, we notice significant differences from what the $\Lambda \mathrm{CDM}$ model predicts [86–96]. These inconsistencies and tensions necessitate exploring alternative theories and modifications to address these issues and refine our understanding expansion of the Universe. 
 
In addressing theoretical and observational challenges, scientists are actively exploring various approaches, including modified gravity theories \citep{Capozziello2002, Carroll2004,Nicolis2009, Deffayet2009,Dvali2000,Nojiri2005,Koivisto2007}, unified DE models \citep{Farnes2018,Cardone2004,Paliathanasis2023,Ansoldi2013,Bousder2020},
interacting DE (IDE) models \citep{Liu2022,Guo2005,Cai2002,Bi2005,Ferreira2017,
 Landim2019,Vagnozzi2020,Costa2014,Kumar2016,Murgia2016,Kumar2017,
Yang2020,Forconi2024,Nunes2021,Yang2018,Lucca2020,
Zhai2023,Bernui2023,Westhuizen2024,Pooya2024,Halder2024,Yao2023,Mishra2023}, 
and the holographic DE (HDE) model \cite{Pavon2005ed,Nojiri2005qp,Wang2005ty,Landim2015fg,Li2009,
Saridakis2017cv,Mamon2017df,Mukherjee2016jk,
 Feng2016hj,Herrera2016gb,Forte2016bb,Nojiri2020cv,Nojiri2021po,Nojiri2021h, Colgain2021s,Wang2016n}, alongside various other alternative cosmological scenarios.

IDE models, which explore the interaction between DE and dark matter (DM), significantly impact the Universe's evolution and the behavior of these mysterious components. 
A phenomenological approach is often used because the exact interaction form is unknown due to the enigmatic nature of DE and DM. Recent observations suggest that a direct DE-DM interaction is plausible.
 
DE can affect the growth of matter perturbations in the Universe through various mechanisms, even without considering interactions between different dark sectors. One way is by slowing down the growth rate due to the Universe's accelerating expansion, which results in a slower evolution of matter perturbations. Moreover, DE can also have perturbations that grow similar to DM, causing alterations in the distribution and clustering of DM in the Universe. These mechanisms illustrate how DE can influence the growth of matter perturbations in the Universe\citep{Pace2010,Garriga1999,Abramo2008,Sapone2012,Basse2014,Nesseris2015,Mota2007, Dossett2013,Batista2013,Batista2014,Ballesteros2008}. In IDE models, DM perturbation growth is influenced by DE perturbations,
 affecting their evolution through energy and possibly momentum exchange between different dark sectors. 
   Researchers use theoretical modeling, simulations, and observational data analysis to understand the growth of perturbations resulting from this interaction    \citep{Westhuizen2024,Sharma2023,Liu2025}.

Although interacting HDE (IHDE) models have been previously studied in the literature, most earlier works were limited in scope—focusing primarily on background data or specific interaction forms, and often neglecting DE perturbations~\cite{Sinha2023,Wang2006,Li2017,Landim2022,Mandal2025,Adhikary2025,
Luciano2023,Nayak2020,Feng2018,Wu2008,Wang2016,Guin2025,Li:2024_DESI}. 
In contrast, the present work provides a more comprehensive and up-to-date investigation of IHDE scenarios. Specifically: (i) we explore three distinct interaction models within a unified HDE framework; (ii) we examine both background expansion and structure growth data, notably including the $f\sigma_8(z)$ dataset; (iii) we analyze the role of DE perturbations, considering both homogeneous and clustered DE, and their impact on the growth of DM perturbations; and (iv) we employ Akaike and Bayesian information criteria (AIC/BIC) to quantitatively compare the models. 

To this end, we first give a brief explanation of the holographic DE model and then derive the coupled differential equations that govern the evolution of DE and DM perturbations in the framework of IHDE models. Solving these equations allows us to directly confront the theoretical predictions with observational data and quantify possible deviations from the standard $\Lambda$CDM model. These aspects clearly distinguish the present study from earlier works and highlight its contribution in providing tighter and more robust constraints on IHDE cosmologies.

The article's structure unfolds as follows:
in Section. \ref{sec2}, we offer a brief overview of the HDE model.
Section. \ref{sec3} is dedicated to deriving the necessary equations governing the background evolution of the Universe and also introduces the IHDE models that are the focus of our study.
Section. \ref{sec4} offers a succinct overview of the current observational datasets at the background level. Using numerical Markov chain Monte Carlo (MCMC) analysis, we then constrain the free parameters within the IHDE models explored in this research.
Section. \ref{sec5} delves into establishing the fundamental equations that govern the evolution of DE and DM within the linear regime of IHDE model scenarios. This section also examines the growth rate of matter perturbations. Furthermore, we merge growth rate data with background data to yield more comprehensive constraints on the free parameters of the models.
In Section. \ref{sec6}, We compare the IHDE model with the $\Lambda$CDM model using cosmological data and parameters. This comparison is based on the best fit parameters obtained from the likelihood analysis.
 Finally, in Sec. \ref{conclude}, we present the conclusions drawn from our study.
\section{HOLOGRAPHIC DARK ENERGY}\label{sec2}
This section explores a DE model derived from the holographic principle, leading to the  HDE model. The holographic principle, introduced by 't Hooft\cite{'tHooft:1993gx} and extended by Susskind\cite{Susskind:1993aa, Susskind:1994vu}, is rooted in the works of Thorn \cite{Thorn:1991fv} and Bekenstein \cite{Bekenstein:1993dz}, forming a basis for quantum gravity and effective quantum field theory. At the Planckian scale, the Universe is described as a two-dimensional lattice evolving in time according to this principle. 
This principle suggests that DE might be explained by vacuum energy, potentially addressing the fine-tuning issue \cite{Cohen:1998zx}. The HDE model proposes that vacuum energy acts as DE, connecting the UV cutoff to the IR cutoff in quantum field theory. 
To prevent the breakdown of the field theory in the presence of gravity, the total energy of a system with size $L$ cannot exceed the mass of a black hole of the same size, expressed as $L^3\Lambda^{4}\leq LM^2_{\rm pl}$ \cite{Cohen:1998zx}, where $\Lambda$ is the UV cutoff and $M_{\rm Pl} = (8\pi G)^{-1/2}$ is the reduced Planck mass. 
The connection between the UV cutoff $\Lambda$ and the IR cutoff $L$ is determined by preventing black hole formation. This criterion necessitates that the enclosed vacuum energy satisfies $\rho_{\rm de}\sim\Lambda^4\leqslant M^{2}_{\rm Pl}L^{-2}$. Li \cite{Li2004} introduced a holographic DE density by considering the entire Universe and interpreting vacuum energy as DE. This model adjusts the relationship between DE density and the future event horizon with a coefficient $ c_{\rm h}$, as a free parameter.
In this scenario, the DE density of the HDE model is given by:
\begin{equation}\label{e1eq1}
\rho_{\rm de}=\frac{{3{c^{2}_{\rm h}} M^2_{\rm pl}}}{L^{2}}
\end{equation}
The choice of IR cutoff parameter $ L$ is crucial in modeling DE, as it determines the scale at which DE density becomes significant.  Some options are considered, each with its merits and demerits.
\begin{itemize}
\item{\bf Hubble horizon:}
 the simplest choice is to set $ L $ equal to the Hubble length ($L = H^{-1}$). This results in a DE density proportional to the square of the Hubble parameter, aligning with observational data and in principle this choice solves the  fine-tuning problem. However, this choice leads to an incorrect equation of state (EoS) for the DE model, failing to explain the observed accelerated expansion of the Universe\citep{Horava2000,Cataldo2001,Thomas2002,Hsu2004}.

\item{\bf Particle horizon:}
using the particle horizon as the IR cutoff also fails to explain the accelerated expansion\cite{Li2004}.

\item{\bf Granda and Oliveros cutoff:}
Granda and Oliveros \cite{Granda2008} propose an alternative length scale $ L $ based on a combination of the Hubble parameter and its time derivative. This approach successfully models the late-time accelerated expansion phase \cite{Granda2008} and aligns with observations of Type Ia supernovae\cite{Wang2017}.

\item{\bf Ricci scale cutoff:}
 the length scale $ L $ can be defined as the curvature of spacetime, specifically the Ricci scalar $ R$ \citep{Gao2009,Zhang2009}. This choice, known as the Ricci HDE model, has been shown to be consistent with observations of Type Ia supernovae\citep{Zhang2009,Easson2011}.

 \item{\bf Event horizon:}
Li's proposal \cite{Li2004} to utilize the event horizon as the IR cutoff for DE is considered the most promising option. This choice, based on the length scale defined by the event horizon, offers a more consistent and accurate model for DE, potentially addressing challenges faced by other options. Moreover, this approach alleviates the coincidence and fine-tuning problems associated with other length scales. 
\end{itemize}
In this work, we adopt the future event horizon, $R_h$, as the IR cutoff.  As defined by Li \cite{Li2004}, the future event horizon is
\begin{equation}\label{eq32}
R_{h}=a\int_{t}^{\infty}\frac{dt}{a(t)}=a\int_{a}^{\infty}\frac{da}{Ha^2(t)},
\end{equation}
where $a$ is the scale factor, $H$ is the Hubble parameter, and $t$ represents cosmic time.  Using this definition, the DE density can be expressed as
\begin{equation}\label{eqroh}
\rho_{\rm de}=\frac{3c^2_{\rm h}M_{\rm Pl}^2}{R_{\rm h}^{2}}\;.
\end{equation}
This approach connects the DE density to the evolution of the Universe by utilizing the future event horizon as the IR cutoff.  Furthermore, as demonstrated in Ref. \cite{Li2004}, Eq.~(\ref{eq32}) yields a DE equation of state (EoS) parameter that closely approaches -1.

Adopting the DE density modeling proposed by Li \cite{Li2004}, which leads to an accelerated Universe and addresses the cosmic coincidence problem when inflation is considered, has been a significant development in this field.

However, extensive studies of the HDE model have revealed that the parameter $ c_{\rm h}$ is consistently less than 1, indicating a phantom Universe with a big rip as its ultimate fate \cite{Zhang2005}. To mitigate this issue, researchers have proposed incorporating a phenomenological interaction between HDE and DM \cite{Li2009,Zhang2012}. This interaction could potentially prevent the big rip by creating an attractor solution where the effective EoS for DE and DM converge in the future.

To further motivate the present study we expand on the rationale for considering IHDE. 
Allowing energy exchange between DE and DM supplies a natural extension with clear physical motivations: such interactions can slow down the relative decline of the matter density, thereby alleviating the coincidence problem, and simultaneously produce distinctive imprints on the background evolution and on linear growth rates that are testable with current redshift-space distortion data. Importantly, IHDE differs from generic IDE models because the interaction is embedded within the holographic setup, retaining a link to quantum-gravity-inspired principles rather than relying exclusively on phenomenological coupling forms\cite{Wang2016,Landim2022,Adhikary2025,Guin2025,Li:2024_DESI,Nayak2020}.
Finally, IHDE scenarios can be mapped, under certain conditions, to effective scalar–tensor or $f(R)$ descriptions, establishing a correspondence with modified gravity frameworks and thereby broadening the theoretical significance of the model \cite{Granda2009,Jawad2013,Gong2009}. This correspondence provides complementary viewpoints for interpreting observational constraints and motivates a thorough study of IHDE both at background and perturbation levels. The present work adopts this theoretically grounded perspective when confronting IHDE with combined background and growth-rate observations.
\section{Basic equations in IHDE models: background level}\label{sec3}
This study examines a homogeneous, isotropic, and flat Universe by utilizing the Friedmann-Robertson-Walker (FRW) metric, which is expressed as \(ds^2 = -dt^2 + a^2(t)(\delta_{ij}dx^idx^j)\), where \(a(t)\) represents the scale factor.
The energy-momentum tensor of DE and DM is denoted as
$  \bar{T}_{\mu\nu} = \bar{p}\bar{g}_{\mu\nu} + (\bar{p} + \bar{\rho})\bar{u}_{\mu}\bar{v}_{\nu} $, where bars represent unperturbed quantities. In the framework of IDE models, the conservation equations for their energy momentum can be written as$\nabla_{\mu}\bar{T}_{\;\;\nu}^{\mu,i} = \bar{Q}^i_{\nu}$ \citep{Marcondes2016}, where, $ i$ represents DE or DM, and $ \bar{Q}^i_{\nu} $ is the phenomenological interaction term between them. Conservation of total energy momentum yields $\bar{Q}^{\mathrm{de}}_{\nu} = -\bar{Q}^{\mathrm{dm}}_{\nu}$. In the absence of interaction, $ \bar{Q}^{\mathrm{de}}_{\nu} = \bar{Q}^{\mathrm{dm}}_{\nu} = 0 $, indicating no energy transfer between DE and DM. Additionally, due to homogeneity and isotropy, the spatial components of $ \bar{Q}^i_{\nu}$ are zero. Thus, the conservation equations for energy density of all components can be expressed as follows:
\begin{align}
&\dot{{\rho}}_{\mathrm{dm}}+3{H}{\rho}_{{\mathrm{dm}}}= Q_{\rm I}  \label{qdw} \\ 
&\dot{{\rho}}_{{\mathrm{de}}}+3{H}(1+w_{\mathrm{de}}){\rho}_{{\mathrm{de}}}=- Q_{\rm I} \label{qsw}\\
&\dot{{\rho}}_{\mathrm{b}}+3{H}{\rho}_{{\mathrm{b}}}=0\label{qhdw} \\ 
&\dot{{\rho}}_{\mathrm{r}}+4{H}{\rho}_{{\mathrm{r}}}=0\label{qldw}  
\end{align}
where dots denote derivative with respect to the cosmic time; $ H$ is the Hubble parameter; $ w_{\mathrm{de}}=p_{\mathrm{de}}/\rho_{\mathrm{de}} $ is the EoS parameter of DE; $\rho_{\rm dm}$, $\rho_{\rm de}$, $\rho_{\rm b}$, and $\rho_{\rm{r}}$ represent the energy densities of DM, DE, baryon, and radiation, respectively; and $Q_{\rm I}$ denotes the phenomenological interaction term. 

The specific expressions of $w_{\mathrm{de}}$ and $Q_{\rm I}$ have a significant impact on the solutions of Eqs. (\ref{qdw} ) and ( \ref{qsw}). This study focuses on the IHDE model, which considers a particular form of $w_{\mathrm{de}}$. Additionally, three phenomenological interaction terms for $Q_{\rm I}$ are examined, describing the energy exchange between DE and DM. Table \ref{tab1222} shows the corresponding interaction terms. 
By investigating these specific expressions, the study aims to understand the behavior and interaction of DE and DM in the IHDE model.

Moreover, the evolution of a spatially flat FRW Universe is described by the equation
 \begin{equation}\label{eqfd}
 H^2=\frac{1}{3M^2_{\rm{pl}}}\big(\rho_{\rm dm}+\rho_{\rm de}+\rho_{\rm b}+\rho_{\rm{r}}\big)
\end{equation}
 By introducing fractional energy densities for different components as $\Omega_{i}={\rho_{i}}/\rho_{\rm cr}$, where $\rho_{\rm cr}=3M_{\mathrm{pl}}^2 H^2$ is the critical density, we can rewrite the Eq. (\ref{eqfd}) as follows:
\begin{equation}\label{eqfr2}
\Omega_{\rm{dm}}+\Omega_{\rm{de}}+\Omega_{\rm b}+\Omega_{\rm{r}}=1.
\end{equation}

By merging Eq. (\ref{eqfd}) with Eqs. (\ref{qdw}--\ref{qldw}), we can obtain the following expression\citep{Feng2016}:
\begin{equation}\label{eqtr}
p_{\rm de}=-\frac{2}{3}\frac{\dot{H}}{H^2}\rho_{\rm cr}-\rho_{\rm cr}-\frac{1}{3}\rho_{\rm r},
\end{equation}
Also, by utilizing Eqs. (\ref{qsw}) and (\ref{eqtr}), we can obtain the following relation:
\begin{equation}\label{eeq56}
2\frac{\dot{H}}{H}(\Omega_{\rm{de}}-1)+\dot{\Omega}_{\rm{de}}+H(3\Omega_{\rm{de}}+\Omega_{\rm I}-3-\Omega_{\rm r})=0,
\end{equation}
where the $ \Omega_{\rm I} $ is defined as $ \Omega_{\rm I}={Q_{\rm I}}/{H\rho_{\rm cr}}$.
Furthermore, using Eqs.~(\ref{eq32}) and (\ref{eqroh}), by definition, 
$R_h^2 = {3 c_{\rm h}^2 M_p^2}/{\rho_{\rm de}}
       = {c_{\rm h}^2}/{\Omega_{\rm de} H^2}$,
we can obtain the following relation:
\begin{equation}
R_{h}=a\int_{a}^{\infty}\frac{d a}{Ha^{2}}=a\int_{x}^{\infty}\frac{d x}{Ha}=\frac{c_{\rm h}}{H\sqrt{\Omega_{\rm de}}}
\end{equation}
 where $x=\ln a$. Hence, we have
 \begin{equation}\label{eeqq12}
 \frac{R_{h}}{a}=\frac{c_{\rm h}}{Ha\sqrt{\Omega_{\rm de}}}
 \end{equation}
Now, by taking the time derivative of both sides of the equation resulting from the combination of Eqs. (\ref{eq32} ) and ( \ref{eeqq12}), the following relation can be obtained:
\begin{equation}\label{ewqry}
\frac{\dot{\Omega}_{\rm{de}}}{2\Omega_{\rm{de}}}+H+\frac{\dot{H}}{H}=\frac{H}{c_{\rm h}}\sqrt{\Omega_{\rm{de}}}.
\end{equation}
The combination of Eqs. (\ref{eeq56} ) and ( \ref{ewqry}) yields the following system of differential equations that govern the background evolution of the Universe in the IHDE model in a flat Universe:
\begin{eqnarray}\label{omepr}
 \Omega^{\prime}_{\rm de}&=&\frac{2}{a}\Omega_{\rm{de}}(1-\Omega_{\rm{de}})\left(\frac{\sqrt{\Omega_{\rm{ de}}}}{c_{\rm h}}+\frac{1}{2}+\frac{\Omega_{\rm r}-\Omega_{\rm I}}{2(1-\Omega_{\rm{de}})}\right)\\
\frac{H^{\prime}}{H}&=&\frac{\Omega_{\rm{de}}}{a}\left(\frac{1}{c_{\rm h}}\sqrt{\Omega_{\rm{de}}}+\frac{1}{2}+\frac{\Omega_{\rm I}-3-\Omega_{\rm r}}{2\Omega_{\rm{de}}}\right)\label{Ehdepr}
\end{eqnarray}
In the equations given, the prime symbol denotes the derivative with respect to the scale factor $a$. The expression for the fractional density of radiation $\Omega_{\rm{r}}(a)$ is defined as $\Omega_{\rm{r0}}a^{-4}/E(a)^2$, where $E(a) = H(a)/H_0$ is the dimensionless Hubble parameter. Here, $\Omega_{\rm{r0}} = \Omega_{\gamma 0}(1+0.2271N_{\rm{eff}})$, with $N_{\rm{eff}} = 3.046$ and $\Omega_{\gamma 0} = 2.469 \times 10^{-5} \times h^{-2}$ with $ H_{0}=100h$  $\rm km\,s^{-1} Mpc^{-1} $ \citep{Hinshaw2013}.
\subsection{The EoS parameter of the IHDE model}
 Here, we investigate the EoS parameter in the IHDE model. Considering the interaction between DM and HDE,
we introduce the energy density ratio \(r = \frac{\rho_{\mathrm{dm}}}{\rho_{\mathrm{de}}} = \frac{\Omega_{\mathrm{dm}}}{\Omega_{\mathrm{de}}}\) for convenience. By utilizing Eqs. (\ref{qdw} ) and ( \ref{qsw}), we derive a differential equation for $ r $ \cite{Wang2005lpv}:
 \begin{equation}\label{evol-r}
 \dot{r}=3Hrw_{\mathrm{de}}+\frac{(1+r) Q_{\rm I}}{\rho_{\mathrm{de}}}.
\end{equation}
Hence, we can obtain:
\begin{equation}\label{eq32w}
w_{\rm de}=-\frac{a\Omega^{\prime}_{\rm de}}{3\Omega_{\mathrm{de}}(1-\Omega_{\mathrm{de}})}-\frac{ Q_{\rm I}}{3H(1-\Omega_{\mathrm{de}})\rho_{\mathrm{de}}}.
\end{equation}
where the prime is derivative with respect to scale factor, $a$.
By replacing the relation $\Omega_{\rm I}={Q_{\rm I}}/{H\rho_{\rm cr}}$ in Eq. (\ref{omepr}) and substituting the obtained result into Eq. (\ref{eq32w}), one can obtain the following expression for the EoS parameter of the IHDE model:
\begin{equation}\label{eq45IHD}
w_{\mathrm{de}}=-\frac{1}{3}-\frac{2\sqrt{\Omega_{\mathrm{de}}}}{3c_{\rm h}}-\frac{ Q_{\rm I}}{3H\rho_{\mathrm{de}}}-\frac{\Omega_{\rm r}}{3(1-\Omega_{\rm de})}.
\end{equation} 
In this work, we consider the following cases for the interaction term $Q_{\rm I}$:
\begin{subequations}
\begin{align}
Q_{\rm I1} &= 3\xi_{1} H \rho_{\rm de}, \label{20aa} \\ 
Q_{\rm I2} &= 3\xi_{2} H \rho_{\rm dm}, \label{20bb} \\
Q_{\rm I3} &= 3\xi_{3} H (\rho_{\rm de} + \rho_{\rm dm}). \label{20cc}
\end{align}
\end{subequations}

where $\xi_{i}$  are dimensionless coupling parameter describing the strength of interaction between DE and DM. Moreover, in this work, we label these interaction cases as IHDE-I, IHDE-II, and IHDE-III models.

Also, the third term of Eq. (\ref{eq45IHD}), by considering the interaction terms (\ref{20aa}--\ref{20cc}), is determined as follows:
\begin{equation}
\frac{ Q_{\rm I}}{3{H} \rho_{\mathrm{de}}}=
\begin{cases}
\xi_{1}\; \;\;\;\;\;\;\;\;\;\;\;\;\;\;\;\;\;  ; $ for $ \text{\small \;\;\; IHDE-I ;}\\
\xi_{2}(\frac{\Omega_{\mathrm{dm}}}{\Omega_{\mathrm{de}}})\;\; 
\;\;\;\;\;\;\;  ;$ for $\;\;\;\;\text{{\small IHDE-II ;}}\\
\xi_{3}(1+\frac{\Omega_{\mathrm{dm}}}{\Omega_{\mathrm{de}}}) \;\;\; ; $ for $ \;\;\;\;\text{{\small IHDE-III}}.
\end{cases}
\end{equation}
The parameter $\xi_{i}$ is used to indicate the transformation of DE into DM when $\xi_{i}>0$, or the transformation of DM into DE when $\xi_{i}<0$. However, the condition $\xi_{i}<0$ often leads to unphysical consequences, such as a negative dark matter density $\rho_{\rm dm}$ and a dark energy density $\Omega_{\rm{de}}$ exceeding 1 in the distant future.

In recent years, there has been significant research on models with an interaction term proportional to the energy densities ($\rho_{\mathrm{de}}$, $\rho_{\mathrm{dm}}$), or a combination of both \citep{Abdalla2009,Abdalla2010,Cao2011,BWang2016,Li2009}.
\begin{table}
\centering
\caption[Equations of state]{ Phenomenological interaction models which have been investigated in this research.}
 \begin{tabular}{lcccccc}
 \hline
 \hline
 Model  & IHDE-I & IHDE-II  & IHDE-III & \\
\hline
${Q_{\rm I}} $  \;\;\;\;\   &  $3{H}\xi_{1}\rho_{\mathrm{de}}$ & $3{H}\xi_{2}\rho_{\mathrm{dm}}$ & \;\; $3{H}\xi_{3}(\rho_{\mathrm{de}}+\rho_{\mathrm{dm}})
 $ & \\
\hline
$ {Q_{\rm I}}/\rho_{\mathrm{de}} $  \;\;\;\;\  & $3{H}\xi_{1}$ & \;\;\;\; $3{H}\xi_{2}\frac{\Omega_{\mathrm{dm}}}{\Omega_{\mathrm{de}}}$& $3{H}\xi_{3}(1+\frac{\Omega_{\mathrm{dm}}}{\Omega_{\mathrm{de}}})  $& \\
\hline
$ {Q_{\rm I}}/\rho_{\mathrm{dm}} $   \;\;\;\;\  &$3{H}\xi_{1}\frac{\Omega_{\mathrm{de}}}{\Omega_{\mathrm{dm}}}$  & \;\;\;\;$3{H}\xi_{2}$   & $3{H}\xi_{3}(\frac{\Omega_{\mathrm{de}}}{\Omega_{\mathrm{dm}}}+1)  $& \\
\hline
\hline
\label{tab1222}
\end{tabular}
\end{table}
\section{OBSERVATIONAL DATA AND METHODOLOGY}\label{sec4}
Here, we present an overview of the steps involved in the analysis of observational data. The analysis comprises the following steps:

\textbf{i.)} Background data analysis: the background level evaluation involves calculating the total likelihood using the $\chi^2_{\text{bac}}$ equation, which combines contributions from various observational datasets, as follows:
\begin{equation}\label{xi22}
 \chi^2_{\text{bac}}(\mathbf{p}) = \chi^2_{\text{SN}}(\mathbf{p}) + \chi^2_{\text{CMB}}(\mathbf{p}) + \chi^2_{\text{H}}(\mathbf{p}) + \chi^2_{\text{BAO}}(\mathbf{p}) 
 \end{equation}
 where $\mathbf{p}$ represents a vector of free parameters for the models, namely $\mathbf{p}=\{\Omega_{b}h^{2}, \Omega_{c}h^{2}, H_{0}, c, \xi_{1}, \xi_{2}, \xi_{3}\}$. The subscripts SN, CMB, $H$, and BAO correspond to the contributions from the supernova type Ia, cosmic microwave background, Hubble parameter, and baryon acoustic oscillations, respectively. The analysis utilizes a total of 1748 observational datasets, including data from the Pantheon+ catalog for SnIa (1701 data points), CMB (3 data points), BAO (8 data points), and Hubble parameter measurements (36 data points).
In the present analysis we fix the primordial perturbation parameters and the reionization optical depth to the Planck 2018 best-fit values from the TT,TE,EE+lowE likelihood, namely
\(\ln(10^{10}A_s)=3.0448\), \(n_s=0.9665\), and \(\tau=0.0544\) \cite{Aghanim et al2020}. This choice is motivated by the very tight constraints that primary CMB anisotropies impose on these quantities, which are substantially stronger than those achievable with the late-time datasets used here. To ensure that late-time growth information is not artificially fixed, we allow the present-day fluctuation amplitude \(\sigma_8\) to vary freely in all our fits; consequently most of the influence of the primordial amplitude on structure growth is absorbed by the fitted \(\sigma_8\). The optical depth \(\tau\) primarily affects large-angle CMB polarization and has only a marginal influence on the background expansion and linear growth observables employed in this work.

\textbf{ii.)} Growth rate data analysis: the analysis of the growth rate incorporates additional data and involves the calculation of the $\chi^2_{\text{tot}}$ value by jointly combining the background and growth rate datasets, as follows: 
\begin{equation}\label{xi222}
     \chi^2_{\text{tot}}(\mathbf{q}) = \chi^2_{\text{bac}}(\mathbf{p}) + \chi^2_{\text{growth}}
     \end{equation}
The models at the background and perturbation levels are characterized by the free parameters $\mathbf{q} = \{\mathbf{p}, \sigma_{8,0}\}$. Additionally, 44 extra data points for the growth rate are integrated with the background data points. For more details about $\chi^{2}_{\text{growth}}$, refer to Sec \ref{growth}.

Statistical tools such as the $\chi^2$ statistic and MCMC techniques are used to assess agreement between theoretical models and observational data, explore parameter space, and determine uncertainties and correlations. These tools analyze observational data, providing constraints on model parameters based on compatibility with the observational data.

\subsection{Type Ia supernovae dataset}
Type Ia supernovae, known for their consistent luminosity, are used as standard candles for cosmological research.
This dataset is fundamental for investigating the Universe's dynamics and constraining DE models by comparing apparent and absolute magnitudes through the distance modulus equation $  \mu_{th}(z)=5\log_{10}{d_{L}(z)} + 42.384 - 5\log _{10}\mathrm{h}$,  where the luminosity distance $ d_{L}(z) $ is defined as $ d_{L}(z)=\frac{c}{H_{0}}(1+z)\int_{0}^{z}\frac{dz^{\prime}}{E(z^{\prime})}$. 
Where $ E(z) $ is the normalized Hubble parameter and defined as: $ E(z) \equiv H(z)/H_{0}$.

The study utilizes the  PantheonPlus dataset containing 1701 data points 
spanning the redshift range $0.01 \leqslant z \leqslant 2.26$
~\cite{Brout2022}. To evaluate the compatibility between the theoretical and observed distance modulus, $\chi^{2}_{\mathrm{sn}}  $ can be calculated as follows:
\begin{equation}
 \chi^{2}_{\mathrm{sn}}(\textbf{p})=\sum_{i=1}^{1701}\frac{[\mu_{th}(\textbf{p}, z_{i})-\mu_{obs}(z_{i})]^{2}}{\sigma_{\mu, i}^{2}}
 \end{equation}
where $  \mu_{th}(\textbf{p}, z_{i}) $ is the theoretical prediction of the distance modulus at redshift $ z_{i} $, $  \mu_{obs}(z_{i}) $ is the observed distance modulus, and $ \sigma_{\mu, i}$ is the uncertainty associated with the observational data.

\subsection{Baryon acoustic oscillations dataset}
Recent studies emphasize the importance of BAO as a powerful geometric tool for probing DE. The position of the BAO peak in the CMB power spectrum is determined by the ratio of the distance scale, $D_{V}(z)$, to the comoving sound horizon, $r_s(z_d)$, at the drag epoch ($z_d$), when baryons decoupled from photons. Komatsu \textit{et al.} \citep{Komatsu2008} pointed out that the drag epoch occurs slightly after photon decoupling ($z_*$), with a larger sound horizon size due to the influence of gravitational potential wells on baryons.
Various studies have measured the BAO feature using different quantities, often focusing on the ratio $r_s(z_d)/D_V(z)$ or its inverse. The comoving sound horizon at the drag epoch, $r_s(z_d)$, is expressed as \citep{Komatsu2008}
\begin{equation}\label{rsa}
r_{s}(z_{d})=\frac{c}{H_{0}}\int^{\infty}_{z_{d}}\frac{c_{s}(z^{\prime})dz^{\prime}}{E(z^{\prime})}
\end{equation}
where $ c_{s}(z)={1}/[{{3}(1+\frac{3\Omega_{b0}}{4(1+z)\Omega_{\gamma 0}})}]^{\frac{1}{2}} $ is the sound speed.
We use an approximation for $z_d$ from \citep{EisensteinHu1998}, and set $\Omega_{\gamma0} = 2.469 \times 10^{-5} h^{-2}$ (for $T_{\mathrm{cmb}} = 2.725 \, \mathrm{K}$) following \citep{Hinshaw2013, Komatsu2008}. The distance scale $D_V(z)$ is given by Ref. \citep{Komatsu2008} as
\begin{equation}
D_{V}(z)=\frac{c}{H_{0}}\Big[(1+z)^{2}D_{A}^{2}(z)\frac{z}{E(z)}\Big]^{\frac{1}{3}}
\end{equation}
where $D_A(z)$ is the angular diameter distance. For a flat Universe ($\Omega_K = 0$), $D_A(z)$ can be calculated as \citep{Komatsu2008}
\begin{align}\label{daaa}
&D_{A}(z)=\frac{c}{H_{0}(1+z)}\int_{0}^{z}\frac{dz}{E(z)}
\end{align}
We analyze two datasets: one in the old format presented in Table \ref{tab1bao} and the other in the new format shown in Table \ref{tab2bao}. Since the data points in these tables are uncorrelated, the $\chi^2$ for the first dataset is calculated as
\begin{equation}\label{bao1}
\chi^{2}_{\mathrm{BAO,1}}=\sum_{i=1}^{4}\frac{[d_{z}(z_{i})|_{th}-(d_{z,i})|_{obs}]^{2}}{\sigma_{i}^{2}}
\end{equation}
where the theoretical prediction is $d_z(z) = \frac{r_s(z_d)}{D_V(z_{\mathrm{eff}})}$, and $D_{V}(z_{\mathrm{eff}})$ denotes the effective volume distance. In the second dataset, $\chi^2_{\mathrm{bao,2}}$ is given by
\begin{equation}\label{bao2}
\chi^{2}_{\mathrm{BAO,2}}=\sum_{i=1}^{4}\frac{[\beta^{*}(z_{i})|_{th}-\beta^{*}_{z,i}|_{obs}]^{2}}{\sigma_{i}^{2}}
\end{equation}
where the theoretical prediction is $\beta^*(z) = \frac{D_V(z_{\mathrm{eff}})}{r_s(z_d)} r_s^{\mathrm{fid}}$. 
\begin{table}
 \centering
 \caption[Equations of state]{Old format of the BAO data points, along with their survey details and references.}
\begin{tabular}{lccccc} 
\hline
\hline
 $z_{\mathrm{eff}}$ & $d_z$ & Survey & Refs. \\
\hline\hline
 0.106& $0.336 \pm0.015$&$\mathrm{6dFGS} $ & \citep{Beutler2011}\\
 \hline
0.60&$ 0.0692\pm0.0033  $&$ \mathrm{WiggleZ} $& \citep{Blake2011} \\
 \hline
0.57&      $ 0.073 \pm.022 $   &   $\mathrm{SDSS.DR9} $  & \citep{Anderson2012} \\
 \hline
0.275&$  0.1390\pm0.0037 $&$\mathrm{SDSS.DR7} $&\citep{Percival2010} \\
\hline
 \label{tab1bao}
 \end{tabular}
 \begin{flushleft}
  \vspace{-0.5cm}
  {\small}
 \end{flushleft}
\end{table}
\begin{table}[ht]
\centering
\caption[Equations of state]{New format of the BAO data points, along with their survey details and references.}
\label{tab2bao}
\begin{tabular}{lcccc}
\hline\hline
$z_{\mathrm{eff}}$ & $\beta_i^*\,(\mathrm{Mpc})$ & $r_{\mathrm{s}}^{\mathrm{fid}}$ & Survey & Refs. \\
\hline\hline
0.38 & $1477 \pm 16$ & 147.78 & BOSS-DR12 & \cite{Alam2017} \\
0.51 & $1877 \pm 19$ & 147.78 & BOSS-DR12 & \cite{Alam2017} \\
0.61 & $2140 \pm 22$ & 147.78 & BOSS-DR12 & \cite{Alam2017} \\
0.15 & $664 \pm 25$  & 148.69 & BOSS-MGS  & \cite{Ross2015} \\
\hline\hline
\end{tabular}
\end{table}

\subsection{Cosmic microwave background dataset}
In this study, we employ the compressed \textit{Planck} 2018 likelihood, which encapsulates the dominant background constraints of the CMB via three distance-related quantities: the acoustic angular scale $l_a$, the shift parameter $R$, and the baryon density $\Omega_b h^2$. While this approach marginalizes over the full shape of the CMB power spectra, it is computationally efficient and widely used in late-time cosmological analyses. Furthermore, it remains compatible with other low-redshift probes such as SnIa, BAO, $H(z)$, and growth-rate data, making it a practical and robust choice for parameter inference across multiple dark energy models.

The position of the CMB acoustic peak is a proper tool to constrain DE models, as it depends on the angular diameter distance in dynamical DE scenarios. This peak in the CMB temperature anisotropy power spectrum is governed by three key parameters: $l_a$, $R$, and $\Omega_b h^2$. 
$l_a$ represents the angular scale of the sound horizon at the time of decoupling. It is computed as
\begin{equation}
l_{a}=(1+z_{*})\frac{\pi D_{A}(z_{*})}{r_{s}(z_{*})}
\end{equation}
where, $z_*$ is the redshift at decoupling, determined using a fitting formula from Hu \citep{Hu1996}, and $D_A(z_*)$ and $r_s(z_*)$ represent the angular diameter distance and the comoving sound horizon at $z_*$, respectively. 
$R$ is the shift parameter at decoupling era, and is defined as \citep{Bond1997}
\begin{equation}
R(z_{*})=\frac{1}{c}\sqrt{\Omega_{m0}}H_{0}(1+z_{*})D_{A}(z_{*})
\end{equation}
The shift parameter $R$ represents the ratio of the distance to the surface of last scattering  to the Hubble radius at the decoupling epoch. In other words, it quantifies how much the Universe has expanded between the time of photon decoupling (around redshift $  z_* \sim 1100  $) and the present day.
Chen \textit{et al.} \citep{Chen et al. 2019} compared the full CMB power spectrum analysis with the distance prior method for constraining DE models and found both approaches to be consistent. In this study, we use the combined CMB likelihood (\textit{Planck} 2018 TT, TE, EE + lowE) based on the observed values reported by Chen \textit{et al.} \citep{Chen et al. 2019}: $ X^{\mathrm{obs}}_{i} = \lbrace R, l_a, \Omega_b h^2 \rbrace = \lbrace 1.7493, 301.462, 0.02239 \rbrace $. The $\chi^2_{\mathrm{cmb}}$ statistic is calculated as
\begin{equation}
\chi^{2}_{\mathrm{cmb}}=\Delta X_{i}\Sigma_{ij}^{-1}\Delta X_{i}^{T}
\end{equation}
where $\Delta X_i = \lbrace X^{\mathrm{th}}_i - X_i^{\mathrm{obs}} \rbrace$ represents the difference between theoretical and observed values, and the inverse covariance matrix $\Sigma_{ij}^{-1}$ is given as
\begin{equation*}
\Sigma_{ij}^{-1} =\left(
\begin{matrix}
94392.3971 & -1360.4913 &1664517.2916 \\
-1360.4913 & 161.4349 & 3671.6180 \\
1664517.2916&3671.6180& 79719182.5162
\end{matrix}
\right)
\end{equation*}

\subsection{{Hubble dataset}}
We employ 36 uncorrelated data points of $H(z)$ spanning $0.07 \leq z \leq 2.34$ from Table \ref{tabHdata}. The $\chi^{2}_{H}$ statistic is expressed as:
\begin{equation}
\chi^{2}_{H}(\textbf{p}) = \sum_{i=1}^{36} \frac{[H_{th}(\textbf{p}, z_{i}) - H_{obs}(z_{i})]^{2}}{\sigma^{2}_{i}}
\end{equation}
In this equation, $H_{th}(\textbf{p}, z_{i})$ denotes model predictions at redshift $z_{i}$, while $H_{obs}(z_{i})$ and $\sigma_{i}$ represent measured values and Gaussian errors, respectively.
\begin{table}
\centering
\caption{The $H(z)$ data used in the current analysis (in
units of  $\mathrm{km\;s^{-1}Mpc^{-1}}$). This compilation is partly based on
those of Ref \citep{Moresco:2016mzx}.}
\begin{tabular}{lclc|lclcccc} 
		\hline \hline
$z$ & $H(z)$ & $\sigma_{H}$ & Refs. & $z$ & $H(z)$ & $\sigma_{H}$ & Refs. \\
\hline
		  $0.07$ & $69 $ & 19.6& \citep{Zhang:2012mp} &$0.48$ & $97 $&62 & \citep{Stern:2009ep}       \\

		 $0.09$ & $69$&12 & \citep{Stern:2009ep} & $0.57$ & $96.8 $ & 3.4& \citep{Anderson:2013zyy}  \\

		  $0.12$ & $68.6 $& 26.2 & \citep{Zhang:2012mp} & $0.593$ & $104$& 13 & \citep{Moresco_2012}  \\

		 $0.17$ & $83$ &  8& \citep{Stern:2009ep} & $0.6$ & $87.9$&  6.1 & \citep{Blake:2012pj}  \\
	
		 $0.179$ & $75$&  4 & \citep{Moresco_2012} &  $0.68$ & $92 $& 8 & \citep{Moresco_2012}  \\
	
		 $0.199$ & $75$&  5 & \citep{Moresco_2012} & $0.73$ & $97.3 $& 7 & \citep{Blake:2012pj}  \\
		
		  $0.2$ & $72.9$&  29.6 & \citep{Zhang:2012mp} & $0.781$ & $105 $& 12 & \citep{Moresco_2012}  \\

		  $0.27$ & $77 $& 14 & \citep{Stern:2009ep} & $0.875$ & $125 $& 17 & \citep{Moresco_2012}  \\

		  $0.28$ & $88.8$ & 36.6& \citep{Zhang:2012mp} & $0.88$ & $90$&  40 & \citep{Stern:2009ep}  \\
	
		  $0.35$ & $82.7 $& 8.4 & \citep{Chuang:2012qt} &  $0.9$ & $117$ & 23& \citep{Stern:2009ep}  \\
	
		  $0.352$ & $83 $& 14 & \citep{Moresco_2012} &  $1.037$ & $154$& 20 & \citep{Moresco_2012}  \\

		 $0.3802$ & $83$&  13.5 & \citep{Moresco:2016mzx} &  $1.3$ & $168$& 17 & \citep{Stern:2009ep}  \\
	
		  $0.4$ & $95$ & 17& \citep{Stern:2009ep} & $1.363$ & $160 $& 33.6 & \citep{Moresco:2015cya}  \\
	
		  $0.4004$ & $77$& 10.2 & \citep{Moresco:2016mzx} &  $1.43$ & $177 $& 18 & \citep{Stern:2009ep}  \\
	
		  $0.4247$ & $87.1$ &  11.2& \citep{Moresco:2016mzx} & $1.53$ & $140$& 14 & \citep{Stern:2009ep}  \\

		  $0.44$ & $82.6$&  7.8 & \citep{Blake:2012pj} &  $1.75$ & $202 $& 40 & \citep{Stern:2009ep}  \\
	
		 $0.44497$ & $92.8$& 12.9 & \citep{Moresco:2016mzx} &$1.965$ & $186.5 $& 50.4 & \citep{Moresco:2015cya}  \\

	    $0.4783$ & $80.9 $& 9 & \citep{Moresco:2016mzx} &  $2.34$ & $222$&7 &  \citep{Font-Ribera:2013wce}  \\
			\hline \hline
	\end{tabular}\label{tabHdata}
\end{table}
\section{Basic equations in IHDE models: linear perturbations level}\label{sec5}
The perturbed FRW metric describes the cosmological spacetime geometry with small perturbations from a flat, homogeneous, and isotropic background universe.
In the conformal Newtonian gauge, the metric is expressed as \(\mathrm{d}\mathrm{s}^2 =\mathrm{a}(\eta)^{2}[ - (1+2\psi)d\eta^{2} + (1-2\phi)\delta_{\mathrm{ij}} \mathrm{dx}^\mathrm{i} \mathrm{dx}^\mathrm{j}]\), where \(a(\eta)\) is the scale factor dependent on conformal time \(\eta\), and \(\psi\) and \(\phi\) represent gravitational potential and spatial curvature. The $ (1+2\psi) $ and $(1-2\phi) $ terms modify the temporal and spatial components of the metric, accounting for small perturbations in the spacetime geometry. In the absence of anisotropic stresses, Einstein's gravity theory requires \(\phi\) and \(\psi\) to be equal, but this equality may not hold in modified gravity models. The perturbed conservation equations with perturbed metrics and energy-momentum tensors yield the following evolution equations for the perturbations\citep{Ma and Bertschinger1995,Marcondes2016,Putter2010}: 
 \begin{align} \notag  
&\dot\delta =-\Big[ 3\mathcal{H}\Big(\frac{\delta p}{\delta\rho}-w_{\mathrm{de}} \Big)-\frac{Q_{\rm I}}{\bar{\rho}}\Big]\delta -\left(1+w_{\mathrm{de}}\right)(\theta -3\dot\phi)\\
&\;\;\;\;\; -\frac{\delta{Q_{\rm I}}}{\bar{\rho}},\label{eq:stbh}\\ \notag
&\dot{\theta}=-\big[\mathcal{H}\left(1-3c^{2}_{\mathrm{a}}\right)-\frac{Q_{\rm I}}{\bar{\rho}}\big]\theta +\frac{\delta p}{\delta\rho}\frac{k^{2}\delta}{\left(1+w_{\mathrm{de}}\right)}+k^{2}\phi \\
&\;\;\;\;\;\; + \frac{i k^{i}\delta Q_{{\rm I}i}}{\bar{\rho}(1+w_{\mathrm{de}})},  \label{eq5} 
 \end{align}
where the dot denotes the derivative with respect to the conformal time, $ \eta $, which is related to the physical time, $t $, through the scale factor, $a$,($a d\eta=dt$). The variables $k^{i}$, $ \delta\equiv\delta\rho/\rho $, and $ \theta$ represent the components of the wave vector in Fourier space, density contrast, and divergence of the peculiar velocity, respectively. The parameter $ w_{\mathrm{de}}$ corresponds to the EoS of DE, taking different values depending on whether the perturbations are associated with dust ($ w_{\mathrm{de}}$=0) or DE. And, $\delta Q_{\mu} $ are the perturbations to the exchange of energy momentum in the perturbed conservation equations. Lastly, the parameter $ c_{\mathrm{a}}^{2} $ represents the squared adiabatic sound speed of the DE perturbations, and its definition is 
\begin{equation}
c^{2}_{\mathrm{a}}=w_{\mathrm{de}}-\frac{\dot{w}_{\mathrm{de}}}{3\mathcal{H}\left(1+w_{\mathrm{de}}\right)}\label{E:m66}
\end{equation}
To investigate perturbations of DE, it is useful to introduce an effective sound speed, $ c_{\mathrm{eff}}$, specifically for DE perturbations. This quantity is defined as \citep{Bean et al.2004}
\begin{equation}
\frac{\delta p}{\delta\rho}=c^{2}_{\mathrm{eff}} +3\mathcal{H}\left(1+w_{\mathrm{d}}\right)\left(c^{2}_{\mathrm{eff}}-c^{2}_{\mathrm{a}}\right)   
\frac{\theta}{\delta}\frac{1}{k^{2}}\label{E77}
\end{equation}
Additionally, in this context, the Poisson equation can be expressed as \citep{Lima1997}
\begin{equation}\label{eq:pois}
k^{2}\phi = - 4\pi G a^{2}(\delta\rho +3\delta p)
\end{equation}
where $ \delta\rho = \delta\rho_{\mathrm{dm}}+\delta\rho_{\mathrm{de}} $ and  $ \delta p = \delta p_{\mathrm{dm}}+\delta p_{\mathrm{de}} $. After that, using quantities $ \delta p_{\mathrm{dm}}=0 $, $ \delta p_{\mathrm{de}}= c^2_{\mathrm{eff}}\delta\rho_{\mathrm{de}} $, $ \delta\rho_{\mathrm{de}}=\rho_{\mathrm{de}}\delta_{\mathrm{de}} $, and $ \delta\rho_{\mathrm{dm}}=\rho_{\mathrm{dm}}\delta_{\mathrm{dm}} $ in   
Eq. (\ref{eq:pois}), the Poisson equation can be written as
\begin{equation}\label{eq:pois2}
- k^{2}\phi =\frac{3}{2}\mathcal{H}^{2}\big[\Omega_{\mathrm{dm}}\delta_{\mathrm{dm} }+(1+3c_{\mathrm{eff}}^2)\Omega_{\mathrm{de}}\delta_{\mathrm{de}}\big]. 
\end{equation} 
In a matter-dominated Universe, the gravitational potential $\phi$ can be considered approximately constant on subhorizon scales $(k^{2}\gg\mathcal{H}^{2})$ in the linear perturbation regime. This assumption is supported by the observation that most structures, which formed during the matter-dominated era, are consistent with this approximation. This simplification facilitates the analysis of perturbation evolution and structure growth. However, this assumption is only valid under certain conditions and may not hold in other regimes or on larger scales \citep{Abramo2009}.

By eliminating $\theta$ from Eqs. (\ref{eq:stbh} ) and ( \ref{eq5}) and using the expressions $\frac{d}{d\eta}=a \mathcal{H}\frac{d}{da}$ and $\frac{d^{2}}{d\eta^{2}}=(a\mathcal{H}^{2}+a\mathcal{\dot{H}}) \frac{d}{da} +a^{2}\mathcal{H}^{2}
\frac{d^{2}}{da^{2}}$, along with Eq. (\ref{eq:pois2}), we can represent Eqs. (\ref{eq:stbh} ) and ( \ref{eq5}) in terms of the scale factor as follows (see Appendix \ref{appendix.A} for more details.)
\begin{align}\label{ewq2}
& {\delta}^{ \prime \prime}_{\mathrm{de}}+A_{\mathrm{de}} {\delta}^{ \prime}_{\mathrm{de}}+B_{\mathrm{de}} {\delta}_{\mathrm{de}}=S_{\mathrm{de}}\\
&  {\delta}^{\prime\prime}_{\mathrm{dm}}+{A}_{\mathrm{dm}} {\delta}^{\prime}_{\mathrm{dm}}+ {B}_{\mathrm{dm}} {\delta}_{\mathrm{dm}}= {S}_{\mathrm{dm}} \label{ecq3}
\end{align}
where, the prime denotes the derivative with respect to the scale factor. 
The coefficients $A_{\text{de}}$, $B_{\text{de}}$, and $S_{\text{de}}$ are expressed by the following expressions:
\begin{eqnarray} \notag
{A}_{\mathrm{de}}&=&\frac{3}{a}+\frac{H^{\prime}}{H}+\frac{3}{a}(c^{2}_{\mathrm{a}}-2w_{\mathrm{de}})-\frac{2}{a^{2}H}\frac{ Q_{\rm I}}{{\rho}_{\mathrm{de}}}\\ \notag
{B}_{\mathrm{de}}&=&\dfrac{3}{a}(c^{2}_{\mathrm{eff}}-w_{\mathrm{de}})\Big[\frac{2}{a}+\frac{ H^{\prime}}{H}-\frac{3}{a}(w_{\mathrm{de}}+c^{2}_{\mathrm{eff}}-c^{2}_{\mathrm{a}})\Big]\\ \notag
&+&\frac{k^{2}c^{2}_{\mathrm{eff}}}{a^{4}H^{2}}-\frac{3}{a}w^{\prime}_{\mathrm{de}}+\frac{1}{a^{3}H^{2}}\Big[3\big (2w_{\mathrm{de}}-c^{2}_{\mathrm{a}}\big)-1\Big]\frac{Q_{\rm I}}{{\rho_{\mathrm{de}}}} \\  
&+&\frac{1}{a^{4}H^{2}}\Big(\frac{Q_{\rm I}}{\rho_{\mathrm{de}}}\Big)^{2}
 -\frac{1}{a^{2}H}\frac{d}{da}\Big(\frac{Q_{\rm I}}{\rho_{\mathrm{de}}}\Big)\\ \notag
{S}_{\mathrm{de}}&=&\frac{3}{2a^{2}} \left(1+w_{\mathrm{de}}\right) \Big[\Omega_{\mathrm{dm}}\delta_{\mathrm{dm}}+
\Omega_{\mathrm{de}}\delta_{\mathrm{de}}\left(1+3c_{\mathrm{eff}}^{2}\right)\Big]\notag   \label{err}
\end{eqnarray}
In addition, the coefficients $ {A}_{\mathrm{dm}} $, $ {B}_{\mathrm{dm}} $, and $ {S}_{\mathrm{dm}}$ are defined as follows:
\begin{align} \notag
&{A}_{\mathrm{dm}}=\frac{3}{a}+\frac{H^{\prime}}{H}-\frac{2}{a^{2}H}\frac{Q_{\rm I}}{{\rho}_{\mathrm{dm}}}\\ \notag
&{B}_{\mathrm{dm}}=-\frac{1}{a^{3}H^{2}}\frac{Q_{\rm I}}{{\rho_{\mathrm{dm}}}}  
+\frac{1}{a^{4}H^{2}}\Big(\frac{Q_{\rm I}}{\rho_{\mathrm{dm}}}\Big)^{2}
 -\frac{1}{a^{2}H}\frac{d}{da}\Big(\frac{Q_{\rm I}}{\rho_{\mathrm{dm}}}\Big)\\ 
&{S}_{\mathrm{dm}}=\frac{3}{2a^{2}}\Big[\Omega_{\mathrm{dm}}\delta_{\mathrm{dm}}+\Omega_{\mathrm{de}}\delta_{\mathrm{de}}\Big]    \label{eroo}
\end{align}
where $ Q_{\rm I} $, $\frac{Q_{\rm I}}{\rho_{\mathrm{dm}}}  $, and $  \frac{Q_{\rm I}}{\rho_{\mathrm{de}}}$ for models IHDE-I, IHDE-II, and IHDE-III are summarized in Table \ref{tab1222}.    
By numerically solving the coupled system of Eqs. (\ref{ewq2} ) and ( \ref{ecq3}), along with Eqs. (\ref{omepr} ) and ( \ref{Ehdepr}), from an initial scale factor of $a_{\mathrm{i}}=10^{-3}$ to the current time ($a=1$), and considering the value of $w_{\text{de}}$ from Eq. (\ref{eq45IHD}), we can determine the density contrasts $\delta_{\text{dm}}$ and $\delta_{\text{de}}$, as well as the fractional energy density of DE ($\Omega_{\text{de}}$) and the Hubble parameter ($H$).
Clustered and nonclustered DE impact DM fluctuations through the effective sound speed parameter \(c_{\text{eff}}\). In the context of clustered DE, $\mathrm{c}_{\mathrm{eff}}\simeq 0$, whereas for nonclustered or uniform DE, $\mathrm{c}_{\mathrm{eff}}\simeq 1$. When dealing with nonclustered DE, simplification occurs by assuming $ \delta_{\text{de}} = 0$. This simplification facilitates the determination of the evolution of density contrasts $ \delta_{\text{dm}} $ and $ \delta_{\text{de}} $ as functions of the scale factor via numerical integration with following suitable initial conditions \citep{Batista2013, Abramo2009}:
\begin{align}
&\delta_{\mathrm{dm,i}}=-2\phi_{i}\Big(1+\frac{k^{2}}{3{\mathcal{H}_{i}}^{2}}\Big) \;\;;\;\ \delta^{\prime}_{\mathrm{dm,i}}=-\frac{2}{3}\frac{k^{2}}{\mathcal{H}^{2}_{i}}\phi_{i}\\ \notag &\delta_{\mathrm{de,i}}=(1+w_{\mathrm{di}})\delta_{\mathrm{dm,i}}\\
&\delta_{\mathrm{de,i}}^{\prime}=(1+w_{\mathrm{di}})\delta^{\prime}_{\mathrm{dm,i}}+w^{\prime}_{\mathrm{di}}\delta_{\mathrm{dm,i}}
\end{align}
where $ w_{\rm di}$ means the value of $ w_{\rm de}$ at $ a_{\rm i}$. 
In order to analyze the growth rate of clustering in the linear regime, a value of $k = 0.1 h \, \text{Mpc}^{-1}$ is chosen to ensure the analysis remains within the linear regime. This choice is supported by the assumption that the power spectrum shape derived from galaxy surveys aligns with the linear matter power spectrum for scales $k \leq 0.15 h \, \text{Mpc}^{-1}$ and is consistent with the power-spectrum normalization $\sigma_8$. The specific selection of $\phi_i = -2 \times 10^{-6}$ corresponds to $\delta_{\text{dm}} = 0.08$ at the present time for $k = 0.11 h \, \text{Mpc}^{-1}$\citep{Smith2003,Tegmark2004,Percival2007}. By using this value of $k$, it will be possible to reliably examine the growth rate of clustering in the linear regime based on the numerical results obtained from solving Eqs. (\ref{ewq2}) and (\ref{ecq3}) along with Eqs. (\ref{omepr} ) and (\ref{Ehdepr}) in the following section.
\begin{table}
 \centering
 \caption[fs8]{$f\sigma_{8}(z) $ data points and their references.}
\begin{tabular}{lcc|cccc} 
\hline
\hline
  $z_{i}$ &  $f\sigma_{8}(z_{i}) $& Refs.& $ z $ &  $f\sigma_{8}(z) $& Refs.\\ 
\hline\hline
 0.02& $ 0.428 \pm0.0465$ & \citep{Huterer et al.:217}&0.3& $ 0.407 \pm0.055$ & \citep{Rita2012}\\
 
0.02& $ 0.398 \pm0.065$& \citep{Hudson:Turnbull2013}& 0.02& $ 0.314 \pm0.048$&\citep{Hudson:Turnbull2013}&\\

 0.38& $ 0.477 \pm0.051$&\citep{Florian2017}& 0.51& $ 0.453 \pm0.050$&\citep{Florian2017}&\\

 0.61& $ 0.410 \pm0.044$&\citep{Florian2017}& 0.76& $ 0.440 \pm0.040$ &\citep{Michael2016}&\\

 1.05& $ 0.280 \pm0.080$&\citep{Michael2016}& 0.32& $ 0.427 \pm0.056$ &\citep{Hector2017}&\\

 0.57& $ 0.426 \pm0.029$&\citep{Hector2017}&0.38& $ 0.497 \pm0.045$ &\citep{Shadab2017}&\\

0.51& $ 0.458 \pm0.038$&\citep{Shadab2017}&0.61& $ 0.436 \pm0.034$ &\citep{Shadab2017}&\\

0.31& $ 0.469 \pm0.098$&\citep{Yuting2017}&0.36& $ 0.474 \pm0.097$&\citep{Yuting2017}&\\

0.40& $ 0.473 \pm0.086$&\citep{Yuting2017}&0.44& $ 0.481 \pm0.076$&\citep{Yuting2017}&\\

0.52& $ 0.488 \pm0.065$&\citep{Yuting2017}&0.48& $ 0.482\pm0.067$&\citep{Yuting2017}&\\

0.59& $ 0.481\pm0.066$&\citep{Yuting2017}&0.56& $ 0.482\pm0.067$&\citep{Yuting2017}&\\

0.64& $ 0.486\pm0.070$&\citep{Yuting2017}&0.10 &$ 0.370 \pm0.130$ &\citep{Feix et al2015}&\\
 
 0.15 &$ 0.490 \pm0.145$&\citep{Howlettetal2015}&0.17 &$ 0.510 \pm0.060$&\citep{SongPercival2009}&\\

 0.18 & $ 0.360 \pm0.090$ &\citep{Blake et al:2013}&0.38 & $ 0.440 \pm0.060$&\citep{Blake et al:2013}&\\
 
0.25 & $ 0.3512 \pm0.0583$ &\citep{Samushia et al. 2012}&0.37 & $ 0.4602 \pm0.0378$&\citep{Samushia et al. 2012}&\\

 0.32 & $ 0.384 \pm0.095$&\citep{Sanchez et al. 2014}& 0.59 &$ 0.488 \pm0.060$ &\citep{Chuang et al. 2016}&\\
 
 0.44& $ 0.413 \pm0.080$&\citep{Blake et al. 2012}& 0.60& $ 0.390 \pm0.063$& \citep{Blake et al. 2012}&\\

 0.73 & $ 0.437 \pm0.072$&\citep{Blake et al. 2012}&0.60 & $ 0.550 \pm0.120$ &\citep{Pezzotta et al. 2017}&\\
 
 1.52 & $ 0.420 \pm0.076$&\citep{Hector2018}& 0.50 & $ 0.427 \pm0.043$&\citep{Rita2012}&\\
   
 0.86 & $ 0.400 \pm0.110$&\citep{Pezzotta et al. 2017}&1.40 & $ 0.482 \pm0.116$ &\citep{Okumura et al. 2016} &\\

0.978 & $ 0.379 \pm0.176$ &\citep{Zhao et al2019} &1.23 & $ 0.385 \pm0.099$ &\citep{Zhao et al2019} &\\   

1.526  & $ 0.342 \pm0.070$ &\citep{Zhao et al2019} &1.944  & $0.364 \pm0.106$ &\citep{Zhao et al2019} &\\ 

\hline\hline
 \label{tab1}
 \end{tabular}
 \begin{flushleft}
  \vspace{-0.5cm}
  {\small}
 \end{flushleft}
\end{table}
\subsection{\textbf{Growth rate data}} \label{growth}
The parameter $f\sigma_8$ can be estimated by numerically solving Eqs. (\ref{ewq2}) and (\ref{ecq3}) along with Eqs. (\ref{omepr}) and ( \ref{Ehdepr}). Here, $f$ represents the linear growth rate of matter perturbations as a function of redshift $z$, which describes the evolution of cosmic structures. It is defined as: $ f={d\ln\delta_{dm}}/{d\ln a} $ \citep{Fry1985,Nesseris2017,Arjona2020,Basilakos2013}.
$\sigma_8(z)$ quantifies the rms mass fluctuations within spheres of radius $8 \, \mathrm{Mpch^{-1}}$ \citep{Nesseris2008}, representing matter clustering on large scales. In the linear regime, it can be expressed as $\sigma_8(z) = \sigma_{8,0} \frac{\delta_m(z)}{\delta_m(z=0)}$. For consistent comparisons across cosmological models, $\sigma_{8,0}$ can be rescaled as $\sigma_{8,0} = \frac{\delta_m(z=0)}{\delta_{m,\Lambda}(z=0)} \sigma_{8,\Lambda}$ \citep{Spyros2015}.
The product $f\sigma_8(z)$ provides insights into galaxy density fluctuations ($\delta_g$), which relate to DM perturbations ($\delta_m$) through the bias factor $b = \delta_g / \delta_m$ \citep{Nesseris2017,Michael2016}. Notably, $f\sigma_8(z)$ is independent of this bias factor \citep{SongPercival2009}, making it a reliable tool for differentiating between cosmological models like IHDE.

In IHDE scenarios, the dark sector interaction modifies the evolution of matter perturbations through additional source terms in the continuity and Euler equations. Nevertheless, the definition of the growth rate, $f \equiv d\ln \delta_m / d\ln a$, remains formally valid as a general kinematic quantity, independent of the specific cosmological model. 
In this work, we compute $\delta_m(a)$ by numerically solving the full set of perturbation equations for IHDE (see Eqs.~(\ref{ewq2})–(\ref{ecq3}), where the interaction terms are explicitly incorporated. The resulting growth rate $f(a)$ thus reflects the impact of the interaction on the clustering of matter. Therefore, the derived quantity $f\sigma_8(z)$ corresponds to an effective growth rate consistent with the IHDE framework.
Theoretical predictions of $f\sigma_8(z)$ can be compared with observational data using the $\chi^2_{\mathrm{growth}}$ statistic:
\begin{equation}
\chi^{2}_{\mathrm{growth}} = \sum_{i=1}^{44} \frac{\left[(f\sigma_{8})_{\mathrm{th}}(z_i) - (f\sigma_{8})_{\mathrm{obs}}(z_i)\right]^2}{\sigma_{\mathrm{obs}}^2(z_i)}.
\end{equation}
This comparison is based on 44 observational measurements of $f\sigma_8(z)$, as detailed in Table \ref{tab1}.
\begin{figure*}
\centering
\setlength\fboxsep{0pt}
\setlength\fboxrule{0pt}
\fbox
{\includegraphics[width=8.5cm]{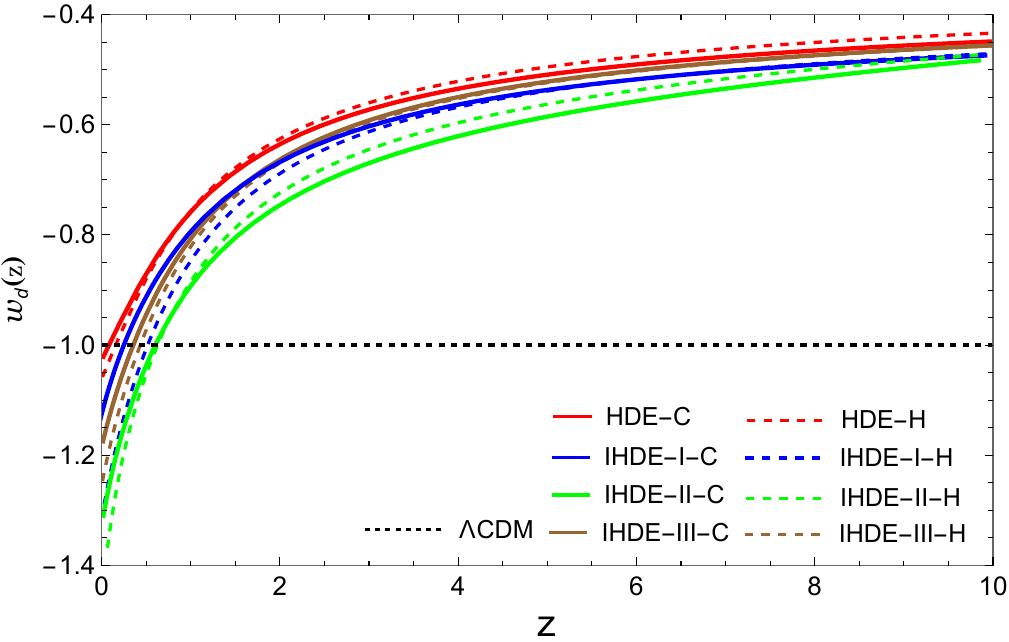}}
{\includegraphics[width=8.6cm]{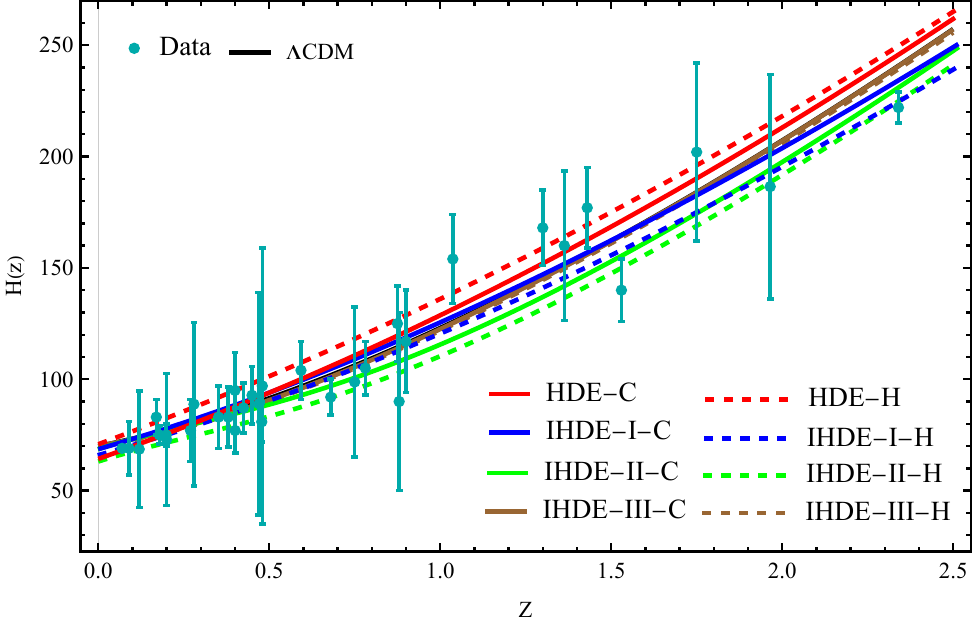}}
{\includegraphics[width=8.5cm]{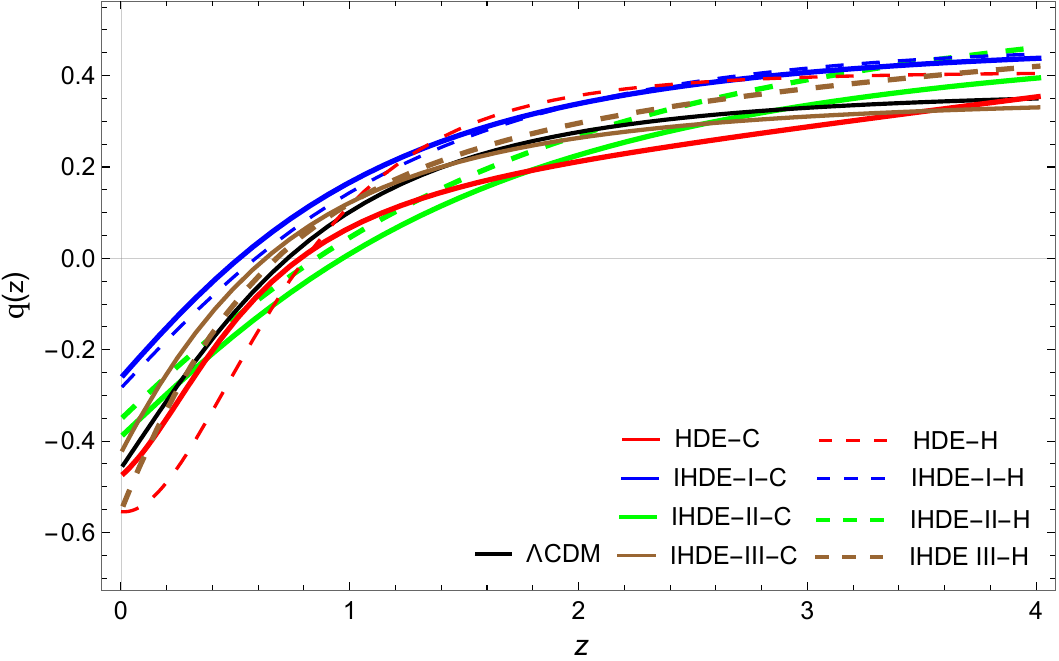}}
{\includegraphics[width=8.5cm]{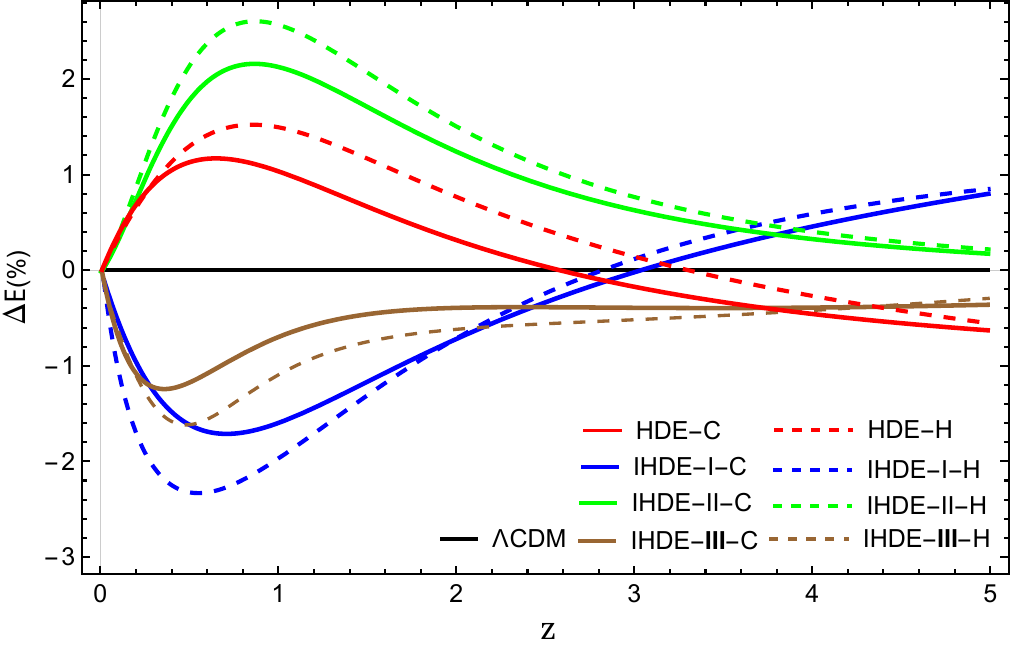}}
\caption{Top left: evolution of the EoS parameter for the HDE and different IHDE  models as a function of redshift z investigated in this work. Bottom left : evolution of the deceleration parameter for various models as a function of $ z$ [see Eq. (\ref{q2zzz})]. Top right: comparison of cosmic chronometer data points from Table \ref{tabHdata} with the theoretical Hubble parameter evolution for HDE and IHDE models with respect to redshift $z.$ Bottom right: $\Delta E(\%) $ of the models compared to the $ \Lambda$CDM model. The various models have been specified by different colors and line styles in the inner panels of the figure. The dashed (solid) line represents
the homogeneous (clustered) case of DE. }
\label{fwd}
\end{figure*}
\begin{figure}
\centering
\setlength\fboxsep{0pt}
\setlength\fboxrule{0pt}
\fbox
{\includegraphics[width=8.6cm]{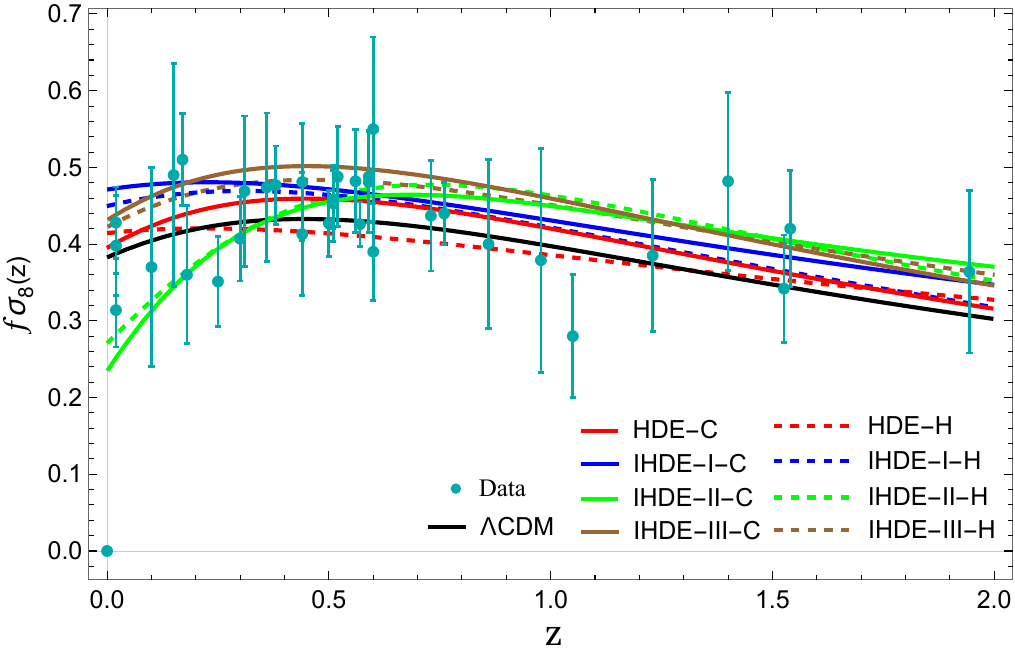}}
{\includegraphics[width=8.6cm]{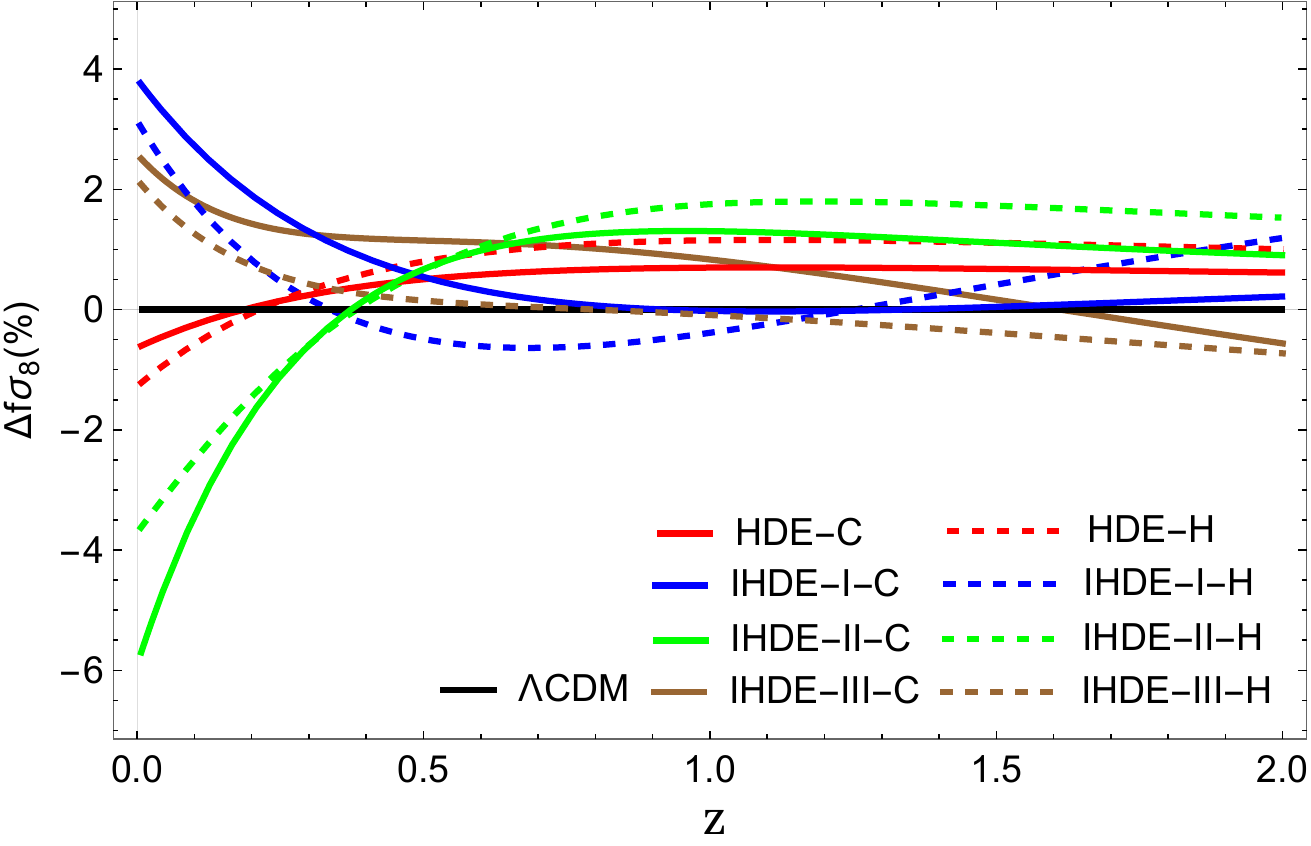}}
\caption{Top: a comparison of observational growth rate data points with the theoretical prediction of the growth rate $f\sigma_{8}(z)$ as a function of redshift $z$ [refer to Sec \ref{growth}].
Bottom: deviations in theoretical predictions of $f(z) \sigma_8(z)$ among $\Lambda$CDM and models in question, normalized to $\Lambda$CDM values [see Eq. (\ref{dels8})].}
\label{fs8}
\end{figure}

\section{ Data Analysis of IHDE Models}\label{sec6}
In this section of the research, we carry out two steps to assess the HDE and IHDE models.

\textbf{i)}: The first step involves applying the MCMC method to constrain the free parameters of the models using the most recent background data [refer to Sec \ref{sec4} and Eq. (\ref{xi22})]. The results of our data analysis on the HDE and IHDE models can be seen in Table \ref{tabbacnu}. 

In the subsequent analysis, we compare and investigate the models by examining key quantities like the EoS parameter, Hubble parameter, deceleration parameter, and Universe's age. This comparative analysis is based on the best-fit parameter values outlined in Table \ref{tabbacnu}. The evolution of these parameters with respect to the redshift $z$ is depicted in Fig. \ref{fwd}.

In the top left panel of the Fig. \ref{fwd}, the evolution of the EoS parameters of the models in terms of $z$ are displayed.
Notably, at redshifts $ z \gtrsim 0.75 $, the EoS parameters for both the homogeneous and clustered DE cases predominantly remain within the quintessence region. However, nearing the present time, except for the HDE model (which nearly reaches the phantom line), all models transition from the quintessence region to the phantom region. Furthermore, the IHDE-II and IHDE-III models cross the phantom line earlier than the other models, occurring at approximately $ z\simeq 0.63 $ and $ z\simeq 0.44 $, respectively.

The Hubble parameter is crucial for understanding the background evolution of the Universe and its impact on the growth of matter perturbations.  Therefore, analyzing its behavior within HDE and IHDE models is important. The bottom right panel of Fig. \ref{fwd} shows the evolution of the percentage deviation of the normalized Hubble parameter,  $\Delta E(\%)$, for various models compared to the standard $\Lambda \mathrm{CDM}$ model. 
This deviation is given by: $\Delta E(\%) = 100 \times \left(\frac{E(z)_{\mathrm{model}}}{E(z)_{\Lambda \mathrm{CDM}}} - 1 \right)$.
 A positive value of $\Delta E(\%)$ indicates that cosmic expansion in the given model is larger than in the $\Lambda \mathrm{CDM}$ model, while a negative value suggests a smaller expansion.
 
The bottom right panel of Fig. \ref{fwd} shows that the IHDE-II( IHDE-III) model consistently display a positive (negative) $\Delta E(\%)$ among all redshifts, $ z $. On the other hand, the HDE (IHDE-I) model present a negative (positive) $\Delta E(\%)$ at high $ z $, which turn positive (negative) in the approximate range of $z\simeq2.54\backsim3.23 $. Furthermore, the $\Delta E(\%)$ for the HDE and IHDE-II (IHDE-I and IHDE-III ) models remains positive (negative), for $ z\lesssim2.54 $. These results hold for both homogeneous and clustered DE related to the models.
The behavior of $\Delta E(\%)$ among the models provides a way to compare how the expansion rate of the Universe differs from $\Lambda \mathrm{CDM}$ model among various epochs. This comparison helps in evaluating the viability of alternative cosmological models and their ability to explain observational data.
Additionally, in the top right panel of Fig. \ref{fwd}, we also compare the theoretical evolution of the Hubble parameter, $H(z)$, with 36 cosmic chronometer data points, which are listed in Table \ref{tabHdata}, which shows that the models are consistent with the observational dataset.
\begin{table*}
\centering
\caption{\label{tabbacnu}
Numerical results of fitting free parameters, with a $1\sigma$ confidence level, obtained for the HDE and IHDE models studied for both homogeneous (H) and clustered (C) dark energy. These results are based on the combination of datasets specified in Eq.~(\ref{xi22}), encompassing background data: BAO + SnIa + CMB + $H(z)$.
}
\centering
\begin{tabular}{lc|ccccccccccc}
\hline
\hline
Model && $\Omega_{\rm b}h^{2}$ & $\Omega_{\rm c}h^{2}$ & $\Omega_{\rm m0}$ & $H_0$ & $ c_{\rm h}$ & $\xi$ & $\chi^2_{\rm min}$ \\ 
\hline
\multirow{2}{*}{HDE}&{\scriptsize H}
& $0.0226^{+0.023}_{-0.025}$ & $0.1280^{+0.012}_{-0.023}$ & $0.3211^{+0.062}_{-0.083}$ & $68.51^{+0.642}_{-0.731}$ & $0.748^{+0.087}_{-0.032}$ & ... & $1714.83^{+2.382}_{-4.241}$ \\ 
&{\scriptsize C}
& $0.0217^{+0.013}_{-0.024}$ & $0.1219^{+0.012}_{-0.023}$ & $0.3082^{+0.042}_{-0.063}$ & $68.28^{+0.765}_{-0.957}$ & $0.798^{+0.062}_{-0.053}$ & ... & $1714.52^{+2.312}_{-3.143}$ \\ 
\addlinespace
\multirow{2}{*}{IHDE-I}&{\scriptsize H}
 & $0.0205^{+0.032}_{-0.041}$ & $0.1238^{+0.012}_{-0.023}$ & $0.3083^{+0.043}_{-0.027}$ & $68.43^{+0.823}_{-0.732}$ & $0.676^{+0.036}_{-0.043}$ & $0.0593^{+0.014}_{-0.016}$ & $1716.14^{+3.112}_{-4.203}$ \\ 
 &{\scriptsize C}
 & $0.0194^{+0.022}_{-0.043}$ & $0.1148^{+0.012}_{-0.023}$ & $0.2981^{+0.052}_{-0.064}$ & $67.13^{+0.626}_{-0.538}$ & $0.752^{+0.039}_{-0.048}$ & $0.0541^{+0.013}_{-0.017}$ & $1715.66^{+2.332}_{-3.145}$ \\ 
 \addlinespace
 \multirow{2}{*}{IHDE-II}&{\scriptsize H}
& $0.0208^{+0.037}_{-0.050}$ & $0.1285^{+0.017}_{-0.020}$ & $0.3113^{+0.057}_{-0.061}$ & $69.25^{+0.747}_{-0.873}$ & $0.617^{+0.071}_{-0.063}$ & $0.0412^{+0.053}_{-0.081}$ & $1715.93^{+3.012}_{-4.323}$ \\ 
 &{\scriptsize C}
& $0.0189^{+0.037}_{-0.039}$ & $0.1110^{+0.017}_{-0.019}$ & $0.2898^{+0.016}_{-0.039}$ & $66.96^{+0.313}_{-0.417}$ & $0.743^{+0.034}_{-0.038}$ & $0.0450^{+0.060}_{-0.052}$ & $1715.45^{+4.011}_{-5.003}$ \\ 
\addlinespace
 \multirow{2}{*}{IHDE-III}&{\scriptsize H}
& $0.0208^{+0.037}_{-0.026}$ & $0.1261^{+0.017}_{-0.020}$ & $0.3098^{+0.057}_{-0.034}$ & $68.84^{+0.446}_{-0.573}$ & $0.686^{+0.067}_{-0.051}$ & $0.0499^{+0.011}_{-0.025}$ & $1715.92^{+4.062}_{-5.073}$ \\ 
&{\scriptsize C}
& $0.0193^{+0.047}_{-0.079}$ & $0.1128^{+0.017}_{-0.019}$ & $0.2939^{+0.077}_{-0.094}$ & $67.04^{+0.610}_{-0.511}$ & $0.730^{+0.057}_{-0.049}$ & $0.0465^{+0.038}_{-0.069}$ & $1715.54^{+3.612}_{-4.227}$ \\ 
\addlinespace
\multirow{1}{*}{$\Lambda\mathrm{CDM}$} &
&{$0.0198^{+ 0.014}_{- 0.016} $  }   & 
{$0.1235 ^{+0.008}_{-0.007} $   }  &
$0.2953^{+0.053}_{-0.067}$ &
{$69.65^{+0.561}_{-0.743} $  } &
{...}                                              &
{$... $  }                                     &
$1714.34^{+2.632}_{-3.049}$\\ 
\hline
\hline
\end{tabular}
\end{table*}
The deceleration parameter is significant in the evaluation of IHDE models, this parameter is defined as
\begin{equation}\label{q2zzz}
q(z) = -\frac{\ddot{a}}{aH^{2}} = \frac{1}{H(z)}\frac{dH(z)}{dz}(1+z) - 1.
\end{equation}
The $ q(z) $ parameter reflects the evolving influence of gravity and DE on the expansion of the Universe.
The deceleration parameter is therefore a key indicator of the Universe's expansion history and the relative strengths of gravity (which causes deceleration) and DE (which causes acceleration).  Its value at different epochs provides valuable insights into the composition and evolution of the Universe. Using Eq. (\ref{q2zzz}) one can determine the redshift $z_{t}$, showing the transition from a decelerating expansion phase of the Universe $  (q > 0) $ to an accelerating expansion phase $ (q < 0) $. Such a transition occurs at $ q = 0 $ or $ \ddot{a} = 0 $.
The Bottom left panel of Fig. \ref{fwd} illustrates the evolution of the deceleration parameter as a function of redshift $z$ for the models examined in this study. The values of the transition redshift $z_t$, considering both homogeneous and clustered DE, are as follows:
{\begin{equation}
z_{t}=
\begin{cases}
 \text{{\small Homogenous DE}}\;\;\;\; \text{{\small Clustered DE}}\;\;\;\;\text{{\small Model}}\\
\;\;\; \;\;\sim  0.79\; ; \;\;\;\;\;\;\;\;\;\;\;\; \;\; \sim 0.77\;\;\;\;\;\;\;\; \;\; \text{{\small HDE ;}}\\
\;\;\; \;\;\sim  0.59\; ; \;\;\;\;\;\;\;\;\;\;\;\; \;\; \sim 0.54\;\;\;\;\;\;\;\; \;\; \text{{\small IHDE-I }};\\
\;\;\; \;\;\sim  0.86\; ;\;\;\;\;\;\;\;\;\;\;\;\; \;\; \sim 0.94\;\;\;\;\;\;\;\; \;\; \text{{\small IHDE-II ;}}\\
\;\;\; \;\;\sim  0.67\; ; \;\;\;\;\;\;\;\;\;\; \;\;\;\;\sim 0.64\;\;\;\;\;\;\;\; \;\; \text{{\small IHDE-III. }}
\end{cases} \notag
\end{equation}}
\begin{figure*}
\centering
\setlength\fboxsep{0pt}
\setlength\fboxrule{0pt}
\fbox
{\includegraphics[width=5.7cm]{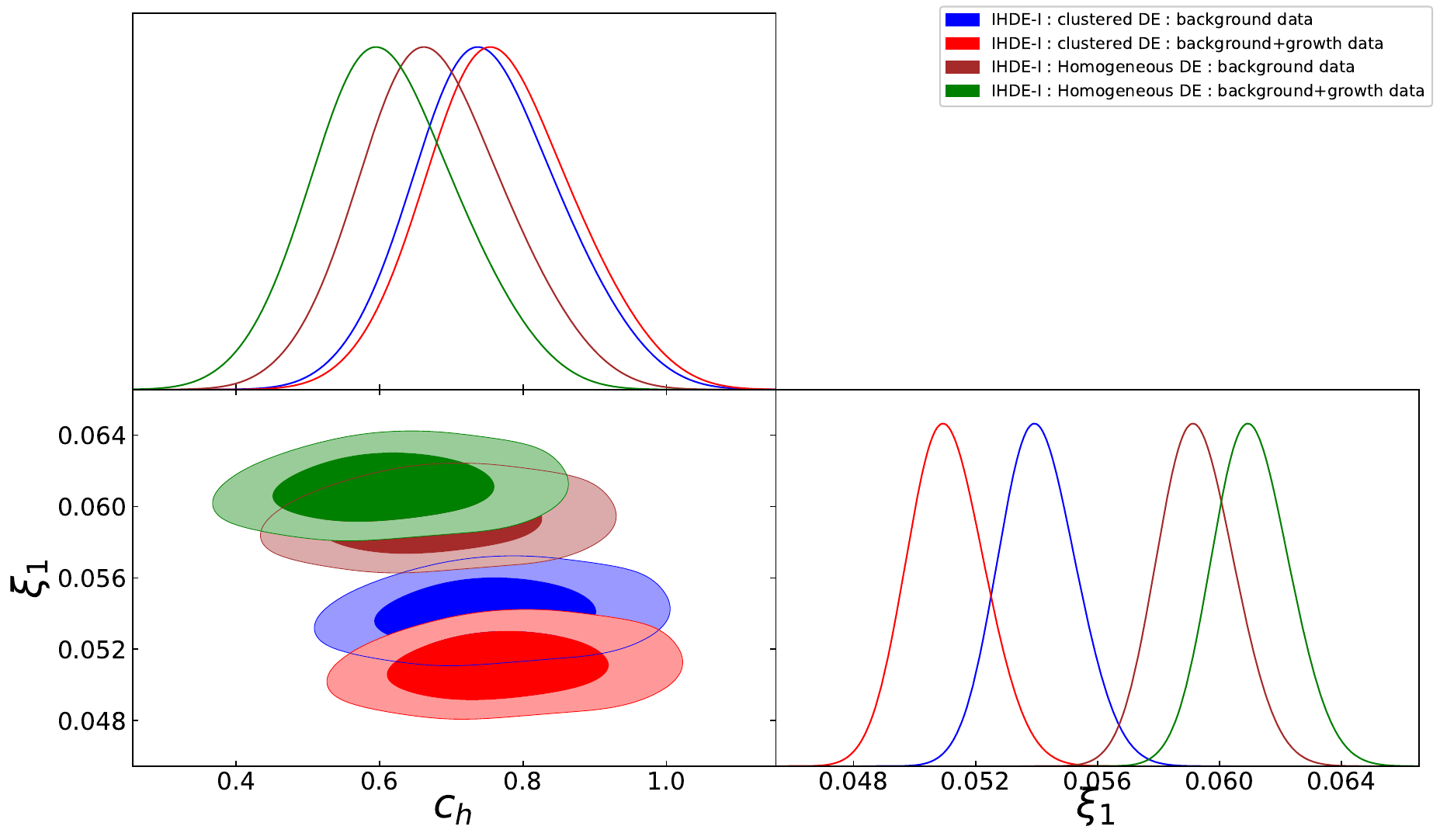}}
{\includegraphics[width=5.7cm]{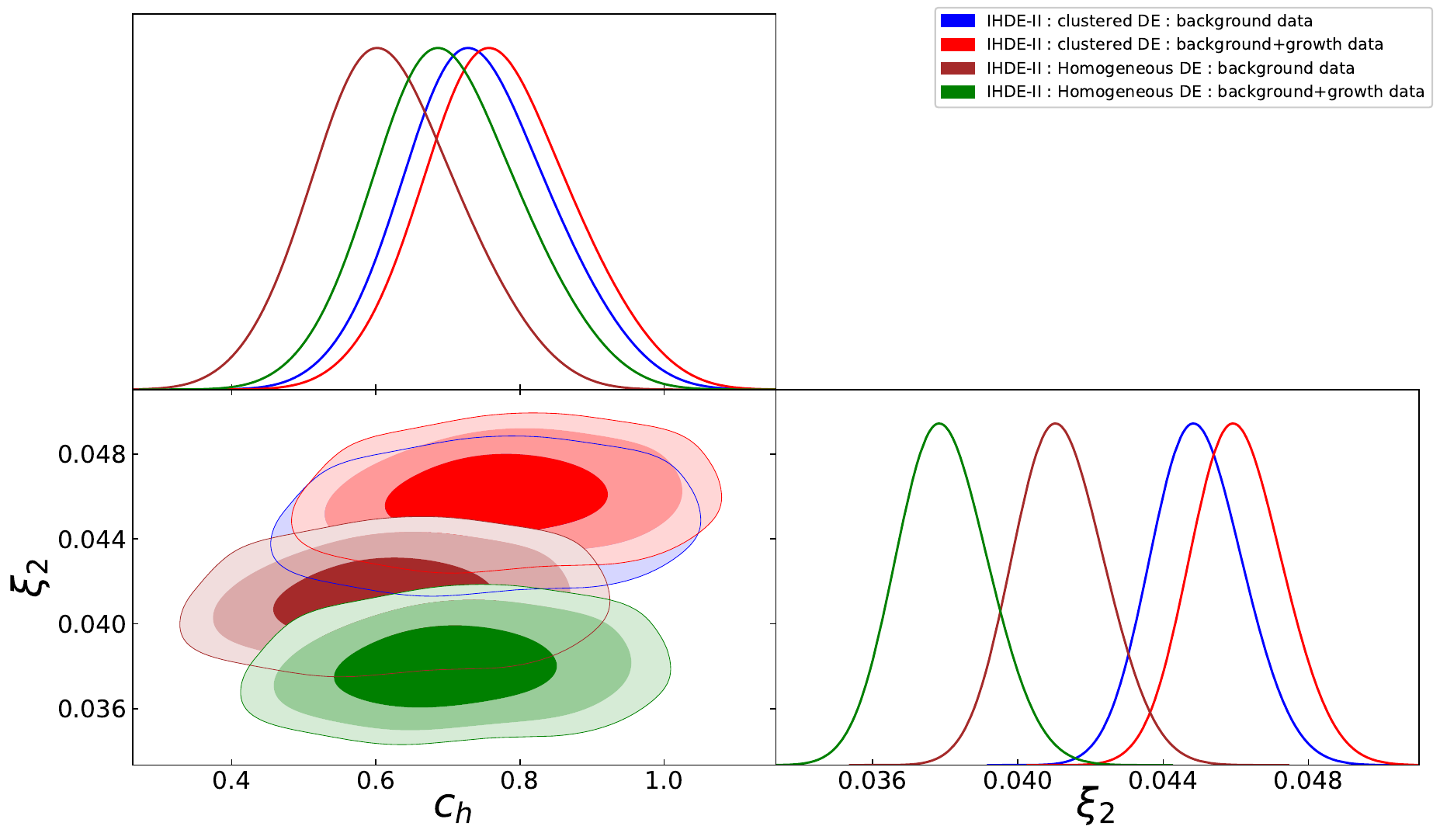}}
{\includegraphics[width=5.7cm]{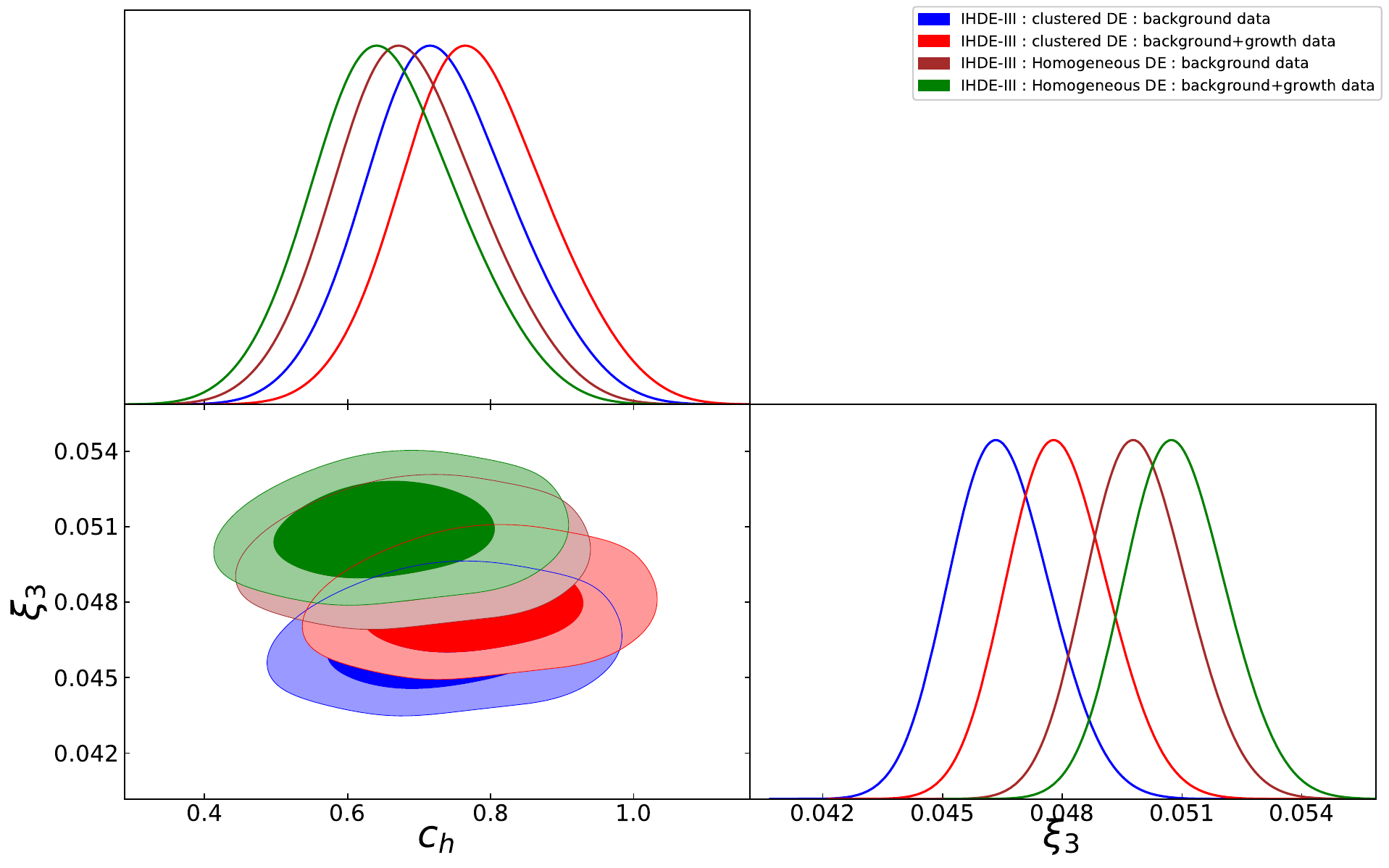}}
\caption{Confidence levels of the $1\sigma$ and $2\sigma$ limits for the IHDE-I (left panel), IHDE-II (middle panel),  and IHDE-III (right panel) models. 
These confidence levels are determined using the background dataset alone for clustered (blue) and homogeneous DE (brown) scenarios. In addition, the combined background and growth rate dataset is utilized for clustered (red) and homogeneous DE (green) scenarios.
}
\label{cxi}
\end{figure*}
Also, in the $\Lambda\mathrm{CDM}$ model, the transition redshift $z_{t}$ is found to be $ z_{t}\simeq 0.711 $. During the early matter-dominated era, the deceleration parameter approaches $\frac{1}{2}$, suggesting a phase of decelerating but slowing expansion. These outcomes agree with the conclusions of the study by Farooq \textit{et al.} in Ref. \citep{Farooq2017}.

\begin{figure*}
\centering
\setlength\fboxsep{0pt}
\setlength\fboxrule{0pt}
\fbox
{\includegraphics[width=8.6cm]{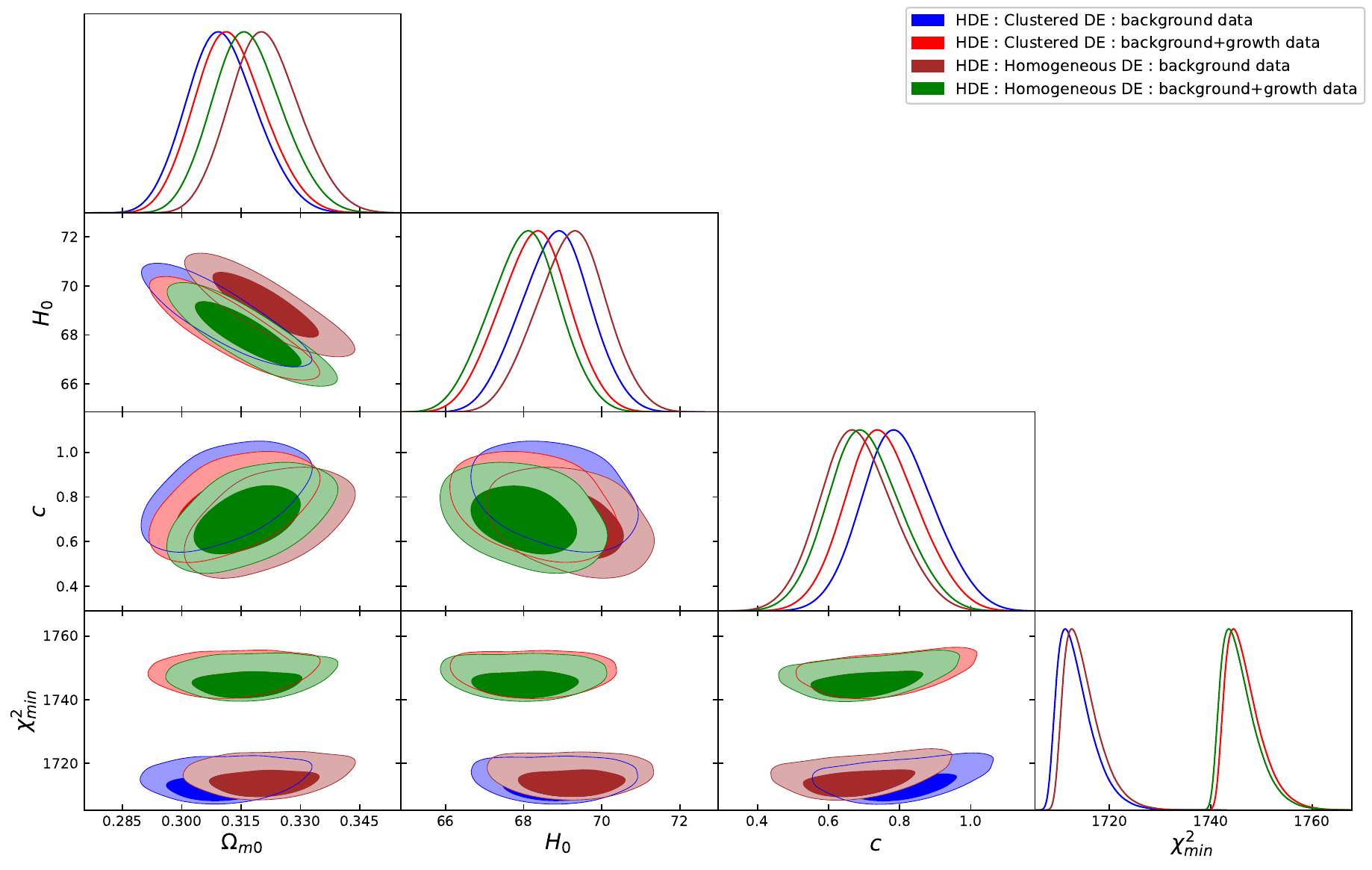}}
{\includegraphics[width=8.6cm]{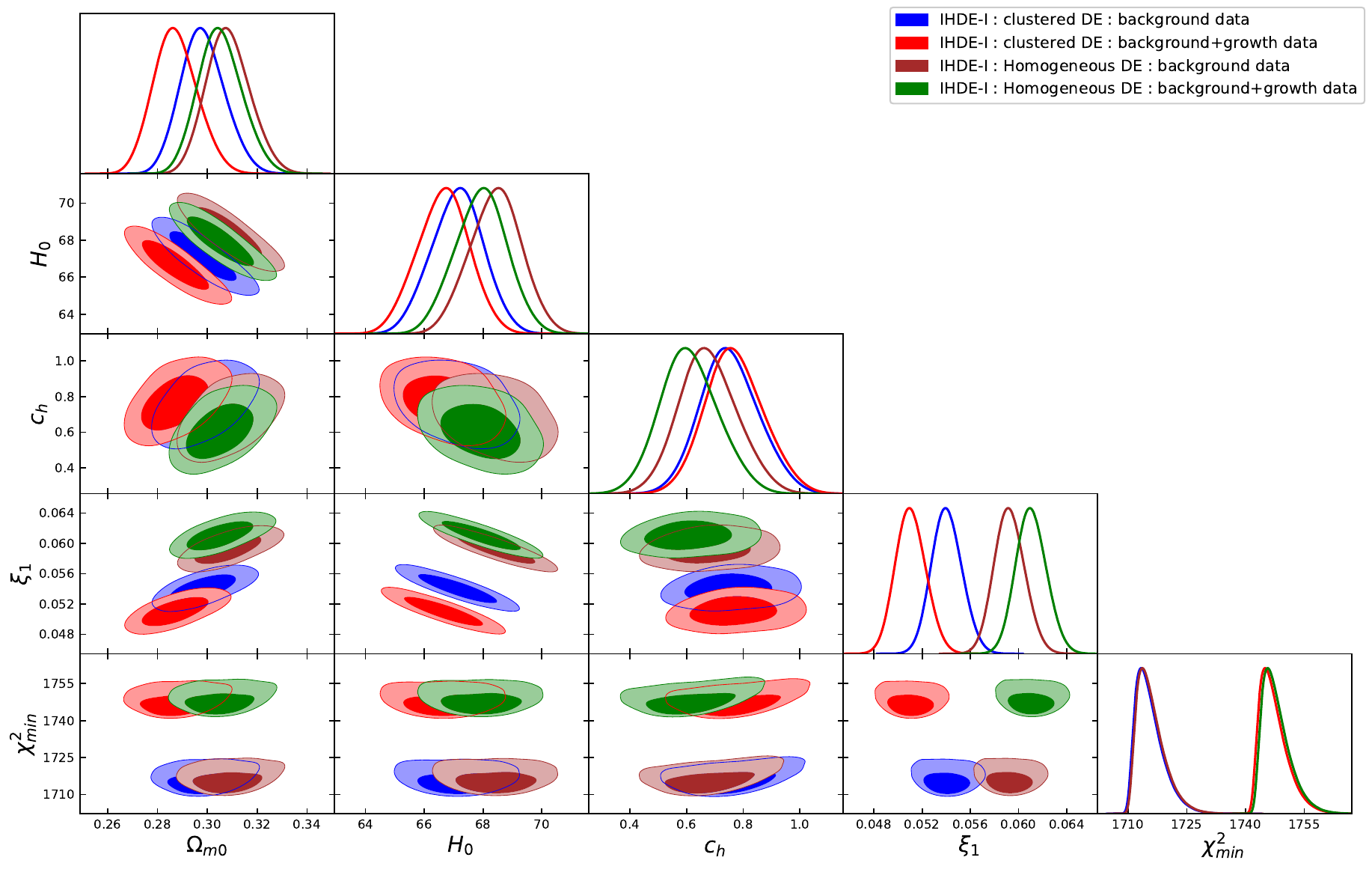}}
{\includegraphics[width=8.6cm]{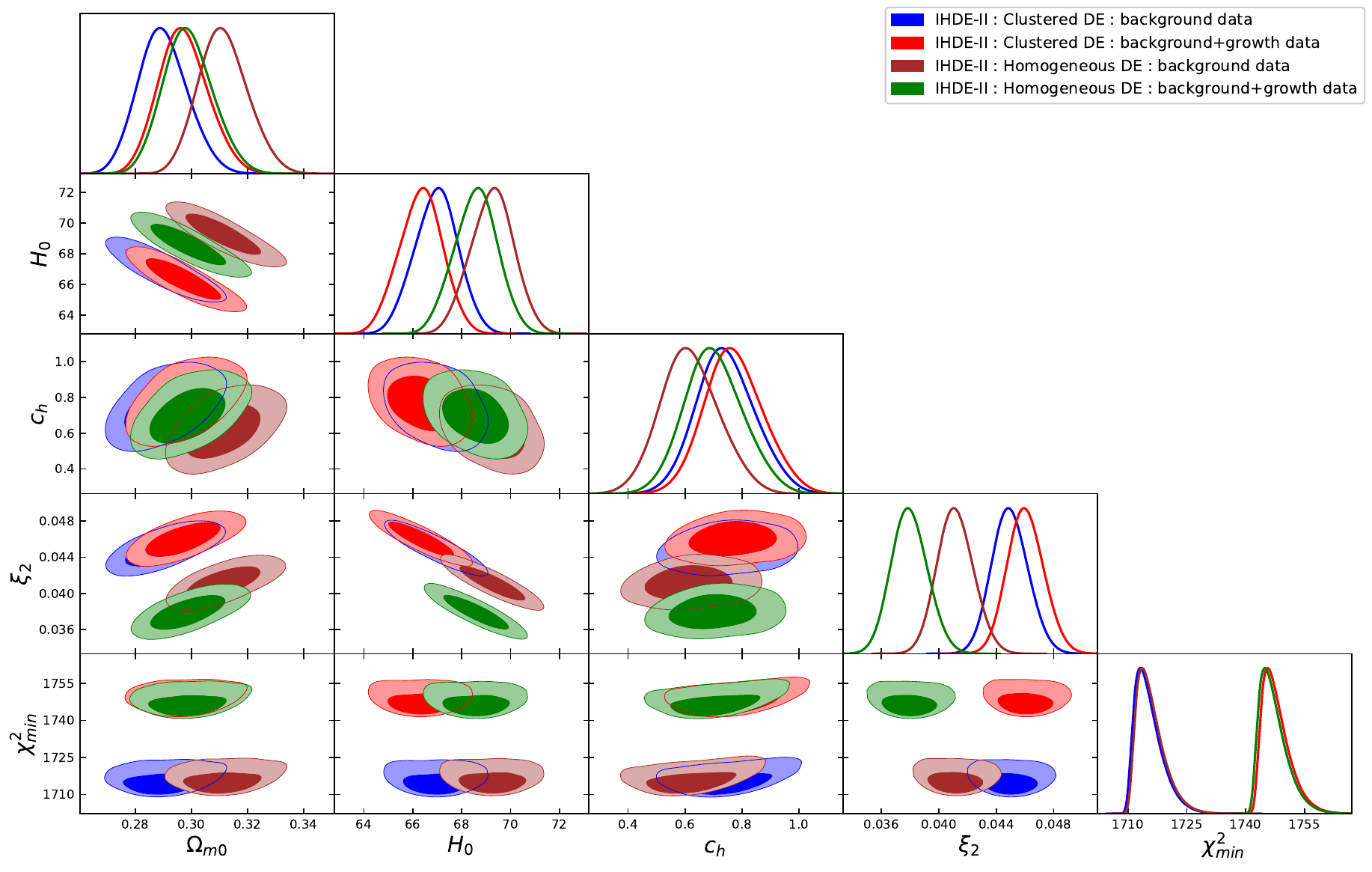}}
{\includegraphics[width=8.6cm]{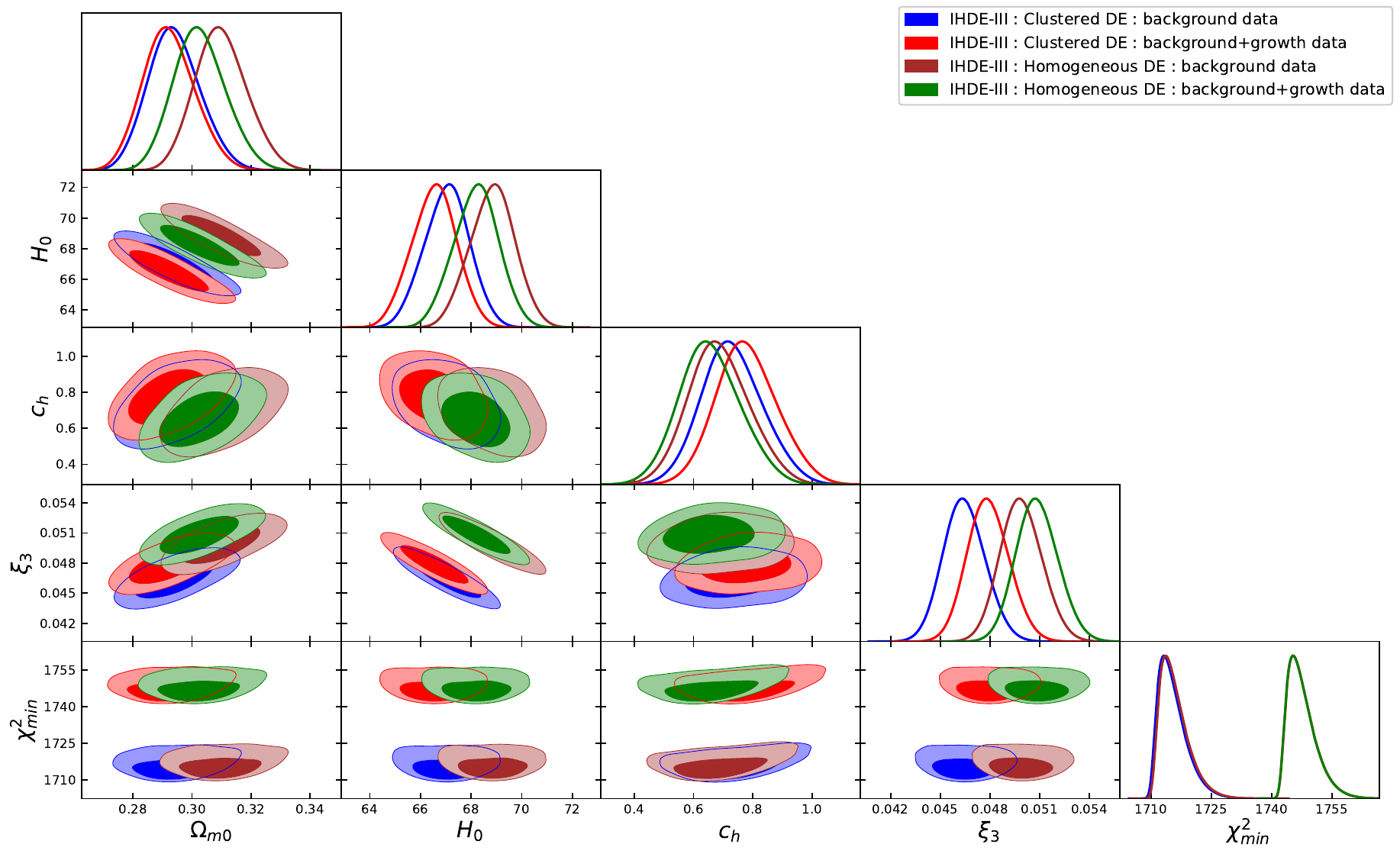}}
\caption{Confidence levels illustrated for the $1\sigma$ and $2\sigma$ limits concerning the HDE and IHDE models. The upper panels display the HDE (left) and IHDE-I (right) models, while the lower panels feature the IHDE-II (left) and IHDE-III (right) models.
These confidence levels have been established using solely the background dataset for both clustered (blue) and homogeneous DE (brown) scenarios. Furthermore, a combined background and growth rate dataset has been employed for both clustered (red) and homogeneous DE (green) scenarios.
For more details, refer to Eqs. (\ref{xi22}) and (\ref{xi222}), along with Tables  \ref{tabbacnu} and \ref{tabclus} for numerical values.}
\label{fr7}
\end{figure*}

\begin{table*}
\centering
\caption{\label{tabclus}Numerical results of fitting free parameters, with a 1$\sigma$ confidence level, for the HDE and IHDE models studied in this research, considering both homogeneous (H) and clustered (C) DE. These results are obtained by combining datasets using Eq. (\ref{xi222}) that include background and growth rate dataset: {$\mathrm{BAO + SnIa + CMB}$ +$H(z)$+ RSD.}}
 \begin{tabular}{lc|cccccccc}
\hline
\hline
Model && $\Omega_{\rm b}h^{2}$ & $\Omega_{\rm c}h^{2}$ & $\Omega_{\rm m0}$ & $H_0$ & $ c_{\rm h}$ & $\xi$ & $\sigma_{8}$ & $\chi^2_{\rm min}$ \\
\hline
\multirow{2}{*}{HDE}&{\scriptsize H}
& $0.0219^{+0.013}_{-0.024}$ & $0.1252^{+0.012}_{-0.023}$ & $0.3152^{+0.042}_{-0.073}$ & $68.33^{+0.542}_{-0.632}$ & $0.733^{+0.067}_{-0.055}$ & ... & $0.821^{+0.082}_{-0.038}$ & $1745.42^{+2.482}_{-4.841}$ \\
 &{\scriptsize C}
&$0.0201^{+0.033}_{-0.054}$ & $0.1254^{+0.012}_{-0.023}$ & $0.3122^{+0.052}_{-0.073}$ & $68.27^{+0.662}_{-0.751}$ & $0.793^{+0.072}_{-0.063}$ & ... & $0.822^{+0.052}_{-0.093}$ & $1746.84^{+3.352}_{-4.163}$ \\
\addlinespace
 \multirow{2}{*}{IHDE-I } &{\scriptsize H}
& $0.0214^{+0.042}_{-0.051}$ & $0.1193^{+0.012}_{-0.023}$ & $0.3051^{+0.063}_{-0.037}$ & $67.93^{+0.522}_{-0.645}$ & $0.610^{+0.052}_{-0.062}$ & $0.0611^{+0.034}_{-0.056}$ & $0.831^{+0.096}_{-0.073}$ & $1747.91^{+3.182}_{-4.063}$ \\
&{\scriptsize C}
& $0.0201^{+0.052}_{-0.063}$ & $0.1074^{+0.012}_{-0.023}$ & $0.2871^{+0.064}_{-0.073}$ & $66.65^{+0.823}_{-0.595}$ & $0.769^{+0.069}_{-0.053}$ & $0.0511^{+0.042}_{-0.041}$ & $0.840^{+0.063}_{-0.057}$ & $1747.32^{+2.673}_{-3.347}$ \\
\addlinespace
 \multirow{2}{*}{IHDE-II }
 &{\scriptsize H}
 & $0.0234^{+0.047}_{-0.062}$ & $0.1172^{+0.017}_{-0.020}$ & $0.2988^{+0.087}_{-0.073}$ & $68.58^{+0.945}_{-0.852}$ & $0.701^{+0.049}_{-0.072}$ & $0.0380^{+0.061}_{-0.073}$ & $0.823^{+0.085}_{-0.0783}$ & $1747.24^{+3.812}_{-3.723}$ \\
 &{\scriptsize C}
& $0.0194^{+0.027}_{-0.046}$ & $0.1113^{+0.017}_{-0.019}$ & $0.2971^{+0.046}_{-0.069}$ & $66.33^{+0.362}_{-0.496}$ & $0.772^{+0.063}_{-0.071}$ & $0.0460^{+0.065}_{-0.058}$ & $0.801^{+0.074}_{-0.088}$ & $1747.91^{+4.011}_{-5.003}$ \\
\addlinespace
 \multirow{2}{*}{IHDE-III }&{\scriptsize H}
& $0.0227^{+0.026}_{-0.051}$ & $0.1181^{+0.017}_{-0.020}$ & $0.3025^{+0.043}_{-0.071}$ & $68.21^{+0.537}_{-0.613}$ & $0.655^{+0.080}_{-0.063}$ & $0.0509^{+0.041}_{-0.035}$ & $0.823^{+0.053}_{-0.072}$ & $1747.62^{+3.462}_{-4.573}$ \\
&{\scriptsize C}
& $0.0198^{+0.065}_{-0.073}$ & $0.1095^{+0.017}_{-0.019}$ & $0.2921^{+0.056}_{-0.083}$ & $66.54^{+0.733}_{-0.828}$ & $0.779^{+0.063}_{-0.071}$ & $0.0479^{+0.029}_{-0.064}$ & $0.824^{+0.077}_{-0.074}$ & $1747.73^{+3.812}_{-4.087}$ \\
\addlinespace
\multirow{1}{*}{$\Lambda\mathrm{CDM}$} &
&{$0.0214^{+ 0.014}_{- 0.016} $  }   & 
{$0.1208 ^{+0.008}_{-0.007} $   }  &
{$0.2913^{+0.053}_{-0.067} $  }  &
{$69.86^{+0.561}_{-0.743} $  } &
{...}                                              &
{$... $  }                                     &
{$0.852^{+ 0.022}_{- 0.022} $ }&$1745.54^{+2.632}_{-3.049}$\\ 
\hline
\hline
\end{tabular}
\end{table*}
Another criterion for assessing the HDE and IHDE models is the age of the Universe, which can be computed using the following equation:
\begin{equation}\label{tu}
t_{_{U}}=\frac{1}{H_{0}}\int_{0}^{\infty}\frac{dz}{(1+z)E(z)}
\end{equation}
This study investigates the age of the Universe ($t_U$) using Eq. (\ref{tu}) for both homogeneous and clustered DE scenarios within the HDE and IHDE models. The best-fitting values from Table \ref{tabbacnu} are used to calculate $t_U$, for homogeneous(clustered), yielding the following results:
HDE: $ 13.578 (13.538)\rm Gyr $,
IHDE-I: $ 13.483 (13.421)\rm Gyr $,
IHDE-II: $ 13.393 (13.356)\rm Gyr $, and within the model  
IHDE-III: $ 13.496 (13.475)\rm Gyr $.
For comparison, the age of the Universe in the standard $\Lambda\mathrm{CDM}$ model is approximately 13.683 Gyr. Notably, the \textit{Planck} 2018 results indicate an age of $13.78 \mathrm{Gyr}$ \citep{Aghanim et al2020}.

\textbf{ii)}: 
In this step of our study, we delve into how the growth of matter perturbations evolves. This task involves the numerical solution of Eqs.~(\ref{ewq2}) and~(\ref{ecq3}) for both homogeneous and clustered DE scenarios within the framework of the HDE and IHDE models. To constrain the ranges of parameters like $\sigma_{8}$ and other parameters related to the HDE and IHDE models, we conduct a unified statistical analysis that merges background and growth rate data obtained from RSD [see Eq. (\ref{xi222}) and Sec \ref{growth}]. The results of these data analysis are outlined in Table \ref{tabclus}.

In the upper panel of Fig.~\ref{fs8}, a comparison is presented between the observed data points (listed in Table \ref{tab1}) and the theoretical prediction of the growth rate of matter perturbations, $f\sigma_{8}(z)$, (refer to Sec \ref{growth}). This analysis includes both homogeneous and clustered DE scenarios within the HDE and IHDE models.

To assess how models affect the growth rate of matter perturbations, we use the following relation:
\begin{equation}\label{dels8}
\Delta f\sigma_{8}(\%)=100\times\Big[\frac{(\Delta f\sigma_{8})_{\mathrm{model}}}{(\Delta f\sigma_{8})_{\mathrm{\Lambda CDM}}}-1\Big]
\end{equation}
By analyzing the sign of $\Delta f\sigma_{8}(\%) $, one can assess how well a specific model aligns with the predictions of the $\Lambda$CDM model regarding the growth rate of matter perturbations in the Universe. A positive (negative) $\Delta f\sigma_{8}(\%) $  indicates that the model in question predicts a greater (lower) growth rate of matter perturbations than the $\Lambda$CDM model.
The values of the $\Delta f\sigma_{8}(\%) $, considering both homogeneous and clustered DE, are as follows:

$ $
{\begin{equation}
\Delta f\sigma_{8}(\%)=
\begin{cases}
 \text{{\small Homogenous DE}}\;\;\; \text{{\small Clustered DE}}\;\;\;\text{{\small Model}}\\
\; \;\;\sim - 0.76\; ; \;\;\;\;\;\;\;\;\;\; \sim - 0.34\;\;\;\;\;\;\; \text{{\small HDE ;}}\\
\; \;\;\sim  +3.06\; ; \;\;\;\;\;\;\;\;\;\;  \sim +4.08\;\;\;\;\;\;\; \text{{\small IHDE-I }};\\
\; \;\;\sim  -3.67\; ;\;\;\;\;\;\;\;\;\;\; \sim -5.41\;\;\;\;\;\;\; \text{{\small IHDE-II ;}}\\
\; \;\;\sim  +2.05\; ; \;\;\;\;\;\;\;\;\;\; \sim +2.47\;\;\;\;\;\;\; \text{{\small IHDE-III. }}
\end{cases} \notag
\end{equation}}
The bottom panel of Fig.~\ref{fs8} reveals that the HDE model most closely tracks the growth rate predicted by the $\Lambda \mathrm{CDM}$ model, and the IHDE-I (IHDE-II) model predicts a faster (slower) growth rate than $\Lambda \mathrm{CDM}$. Furthermore, the impact of DE clustering versus homogeneity on the growth rate $(\Delta f \sigma_8 (\%))$ varies between models. For example, the IHDE-I model predicts a faster growth rate with clustered DE compared to homogeneous DE, while the IHDE-II model shows the opposite behavior.

Although the correlations between $c_{\rm h}$ and the interaction parameters also appear in Fig. \ref{fr7}, we include Fig.~\ref{cxi} separately to improve clarity and emphasize these specific relationships. Presenting these subsets individually facilitates a more focused interpretation of degeneracies that might be visually suppressed in the complete correlation matrix.
\begin{table}
\centering
\caption{Model selection results, using background dataset for homogeneous (H) and clustered (C) DE within the HDE and IHDE models (see Sec. \ref{MSC}).}
  \begin{tabular}{lc|cccccccc}
\hline
\hline
Model &&\;\;$\mathrm{k}$ &$\chi^{2}_{\mathrm{min}}$ 
  &$\mathrm{AIC}$&$ \mathrm{\Delta AIC}$&$ \mathrm{BIC} $&$\Delta  \mathrm{BIC} $&\\ 
\hline
\hline 
\multirow {2}{*}{HDE}
& {\footnotesize H}&\;4 &1715.92 & 1723.92& 3.58 & 1745.78  & 9.04\\ 
& {\footnotesize C}&\;4 & 1715.56 &  1723.56 &3.22  & 1745.42  & 8.68\\
\hline
\multirow {2}{*}{IHDE-I}
& {\footnotesize H}&\;5&1715.45 & 1725.45 & 5.11& 1752.78& 16.04 \\
& {\footnotesize C}&\;5&1715.46 & 1725.46 &5.12  &1752.79 &16.05 \\
\hline
\multirow{2}{*}{IHDE-II}
& {\footnotesize H}&\;5 & 1715.73 & 1725.73&5.39 &1753.06 &16.32 \\
& {\footnotesize C}&\;5 & 1715.93 &1725.93 &5.59 &1753.26 &16.52 \\
\hline
\multirow{2}{*}{IHDE-III}
& {\footnotesize H}&\;5 & 1714.53& 1724.53&4.19 &1751.86  & 15.12  \\
&{\footnotesize  C}&\;5 & 1714.83 &1724.83 & 4.49 & 1752.16 & 15.42 \\
\hline
$\Lambda$CDM&...&\;3 & 1714.34 & 1720.34&0.0&1736.74 &0.0 \\
\hline
\hline
\end{tabular}
\label{tabbac}
\end{table}
Panels of Fig.~\ref{cxi} illustrate the correlations between $c_{\rm h}$ and the $\xi$ parameters $(\xi_1, \xi_2, \xi_3)$ for the IHDE-I, IHDE-II, and IHDE-III models, respectively. These correlations are analyzed under both clustered and homogeneous DE assumptions, using background data alone and background data combined with growth rate data.
The left panel of Fig.~\ref{cxi} shows an inverse correlation between $\xi_{1}$
and $ c_{\rm h}$ in the IHDE-I model when both datasets are used under clustered and homogeneous DE assumptions. Moreover, when solely employing background data, the parameter $\xi_{1}$ ($ c_{\rm h}$) in the clustered DE scenario is smaller (larger) than its value in the homogeneous DE scenario.
The middle panel of Fig. \ref{cxi} demonstrates that in the IHDE-II model, using clustered DE with both datasets results in higher values of $ \xi_{2}$ and $ c_{\rm h}$ compared to using only background data.  Conversely, with homogeneous DE and both datasets, $\xi_{2}$ is lower, but $ c_{\rm h}$ is higher. 
The right panel of Fig. \ref{cxi} shows that in the IHDE-III model, incorporating both datasets and the DE assumption can impact the values and magnitudes of $ c_{\rm h}$ and $ \xi_{3} $.
\begin{table}
\centering
\caption{Model selection results, using background and growth rate data, for homogeneous (H) and clustered (C) DE within the HDE and IHDE models (see Sec. \ref{MSC}).}
  \begin{tabular}{lc|cccccccc}
\hline
\hline
Model &&\;\;$\mathrm{k}$ &$\chi^{2}_{\mathrm{min}}$ 
  
&$\mathrm{AIC}$&$ \mathrm{\Delta AIC}$&$ \mathrm{BIC} $&$\Delta  \mathrm{BIC} $&\\ 
\hline
\hline
\multirow {2}{*}{HDE}
& {\footnotesize H}&\;5 & 1746.84 & 1756.84   &3.30  &1784.30&8.80 \\
& {\footnotesize C}&\;5 & 1746.42 & 1756.42  & 2.88 &1783.87&8.37 \\
\hline
\multirow {2}{*}{IHDE-I}
& {\footnotesize H}&\;6& 1747.24 & 1759.24 &5.70  & 1792.18&16.68 \\

& {\footnotesize C}&\;6 & 1747.32 & 1759.32  & 5.78 &1792.26& 16.76\\
\hline
\multirow{2}{*}{IHDE-II}
& {\footnotesize H}&\;6 & 1747.91 & 1759.91 &6.37  &1792.85&17.35 \\
& {\footnotesize C}&\;6&1748.41 & 1760.41 & 6.87 &1793.35 &17.85 \\
\hline
\multirow{2}{*}{IHDE-III}
&{\footnotesize  H}&\;6 &  1747.62 &  1759.62  &6.08  &1792.57 &17.07 \\
& {\footnotesize C}&\;6 & 1747.73&   1759.73 &6.19  & 1792.66&17.16 \\
\hline
$\Lambda$CDM&...&\;4 & 1745.54 & 1753.54  &0.0& 1775.50  &0.0 \\
\hline
\hline
\end{tabular}
\label{tabbacgr}
\end{table}

Furthermore, panels of Fig. \ref{fr7} display the $1\sigma$ and $2\sigma$ confidence levels for the models, determined using background and  growth rate data for both homogeneous and clustered DE scenarios. These triangular plots visually show the correlations among each pair of model parameters.
\subsection{\textbf{Model selection criteria}}\label{MSC}
When comparing models, we can evaluate their performance using several key criteria. For models with the same number of degrees of freedom, a lower $ \chi $-squared value ($\chi^{2}_{\mathrm{min}}$) indicates a better fit to the data.  For unequal degrees of freedom, the reduced $ \chi $-squared statistic, $\chi^{2}_{\mathrm{red}} = \chi^{2}_{\mathrm{min}} / (N - k)$, where $k$ is the number of free parameters and $N$ is the number of data points, is used. A value near 1 suggests a good fit with observational data; values significantly above or below 1 indicate a poor fit.
The Akaike Information Criterion \citep{Akaike1974} and Bayesian Information Criterion \citep{Schwarz1978} offer alternative approaches.  They are defined as: $\mathrm{AIC} = -2\ln \mathcal{L}_{\mathrm{max}} + 2k+2k(k+1)/(N-k-1)$ and $\mathrm{BIC} = -2\ln \mathcal{L}_{\mathrm{max}} + k\ln N$, where $\mathcal{L}_{\mathrm{max}}$ is the maximum likelihood, related to $\chi^{2}_{\mathrm{min}}$ by $\chi^{2}_{\mathrm{min}} = -2\ln \mathcal{L}_{\mathrm{max}}$.  Here, $N$ is 1748 for background data and 1792 when including growth data (Sec. \ref{growth}).
To compare models, the differences $\Delta\mathrm{AIC}$ and $\Delta\mathrm{BIC}$ relative to a reference model (often the best-fitting model) are calculated \citep{Rivera2019, Liddle2007} as follows:
\begin{align}\label{AIC}
\Delta \mathrm{AIC} &= \mathrm{AIC}_{\mathrm{model}} - \mathrm{AIC}_{\Lambda \mathrm{CDM}} = \Delta \chi^{2}_{\mathrm{min}} + 2\Delta k \\
\Delta \mathrm{BIC} &= \mathrm{BIC}_{\mathrm{model}} - \mathrm{BIC}_{\Lambda \mathrm{CDM}} = \Delta \chi^{2}_{\mathrm{min}} + \Delta k (\ln N) \label{BIC}
\end{align}
Smaller $|\Delta\mathrm{AIC}|$ and $|\Delta\mathrm{BIC}|$ values indicate better model support.  Guidelines from Ref. \citep{Rivera2019} suggest substantial support for $|\Delta \mathrm{AIC}| \in (0, 2]$, less support for $|\Delta \mathrm{AIC}| \in [4, 7]$, and rejection for $|\Delta \mathrm{AIC}| > 10$. Similar interpretations apply to $|\Delta \mathrm{BIC}|$, with larger values suggesting better agreement with the observational data. The choice of criterion depends on the degrees of freedom and the balance between goodness of fit and model complexity.

We illustrate the computed results in Tables \ref{tabbac} and \ref{tabbacgr}, utilizing the numerical values found in Tables \ref{tabbacnu} and \ref{tabclus}. 
Based on the AIC and BIC analysis, it can be concluded that the selection of a model that best fits with the observational data (background vs. background + growth rate) depends on two factors: the DE assumptions (clustered vs. homogeneous)  and the interaction terms between DE and DM. Additionally, the model selection also depends on the specific dataset being employed.

For example, based on AIC analysis of the background data alone, the model ranking in terms of best fit to the observational data is HDE,
IHDE-III, IHDE-I, and IHDE-II, whether it is for homogeneous or clustered DE.

On the other hand, when homogeneity in DE is considered, all models except HDE imply that assuming homogeneity leads to a slightly better agreement with observational data compared to considering clustered DE.
Overall, this assessment suggests that in scenarios involving homogeneous DE, the IHDE models exhibit a slightly better alignment with observational data compared to models that assume clustering for DE (see Table \ref{tabbac}).

Additionally, when conducting AIC analysis using both background and growth rate data, it can be concluded that the HDE and IHDE-I models fit the observational data better than other models. This holds true for both homogeneous and clustered DE. Specifically, for clustered DE, the HDE model demonstrates a slightly better consistency compared to homogeneous DE. The IHDE-III and IHDE-II models follow next in terms of compatibility with the observational data. However, the assumptions regarding homogeneity and clustering of DE can alter the order of how well the models fit the data (see Table \ref{tabbacgr}).
\section{DISCUSSIONS AND CONCLUSIONS}\label{conclude}
In this study, we employed a two-step approach to analyze HDE and IHDE models with three different DE-DM interaction terms. Initially, we conducted an MCMC analysis to constrain model parameters using the latest background data (see Sec. \ref{sec4} and Eq. (\ref{xi22})). The resulting constraints for the HDE and IHDE models are summarized in Table \ref{tabbacnu}.  

In the top left panel of Fig. \ref{fwd}, we depicted the evolution of the EoS parameters in terms of redshift $z$.
 Notably, at redshifts $z \gtrsim 0.75$, EoS parameters for both DE scenarios primarily remained in the quintessence region. However, as we neared the present time, all models, except for the HDE model (which approached the phantom line), transitioned from quintessence to the phantom region. The IHDE-II and IHDE-III models crossed the phantom line earlier, at approximately $z \simeq 0.63$ and $z \simeq 0.44$, respectively.

In the bottom right panel of Fig. \ref{fwd}, we illustrated the evolution of the percentage deviation of the normalized Hubble parameter, $\Delta E(\%)$, comparing various models with the standard $\Lambda \mathrm{CDM}$ model. A positive (negative) $\Delta E(\%)$ implies a higher (lower) expansion rate than $\Lambda$CDM. 
The IHDE-II (IHDE-III) model consistently showed a positive (negative) $\Delta E(\%)$ across all redshifts, while the HDE (IHDE-I) model displayed a negative (positive) $\Delta E(\%)$ at high $z$, switching signs around $z \simeq 2.54 \sim 3.23$. The behavior of $\Delta E(\%)$ across models provided insights into the expansion rate of the Universe relative to the $\Lambda \mathrm{CDM}$ model.
In the top right panel of Fig.~\ref{fwd}, we compared the theoretical evolution of the Hubble parameter, $H(z)$, with cosmic chronometer data listed in Table~\ref{tabHdata}, confirming consistency between the models and observational data.
We found that the transition time from decelerated to accelerated expansion in the studied IHDE models was comparable to that in the $\Lambda$CDM model. The transition time of the HDE and IHDE-III models was particularly close to that of the $\Lambda$CDM model in both DE scenarios. Additionally, we calculated the age of the Universe within each IHDE model, observing that the ages in the HDE and IHDE-III models were better aligned with the standard $\Lambda$CDM model for both homogeneous and clustered DE scenarios.

Next, we shifted our focus to the growth of matter perturbations within the HDE and IHDE frameworks, solving Eqs. (\ref{ewq2}), (\ref{ecq3}), (\ref{omepr}), and (\ref{Ehdepr}) for both DE scenarios. We constrained the parameters of the HDE and IHDE models, including $\sigma_8$, through a comprehensive statistical analysis using background and growth rate data. 

Correlations between the HDE parameter $c_{\rm h}$ and the interaction parameters $(\xi_{1}, \xi_{2}, \xi_{3})$ for the IHDE models are shown in Fig. \ref{cxi}, under both clustered and homogeneous DE assumptions. 
When both background and growth data were used, an inverse correlation between $\xi_1$ and $c_{\rm h}$ was observed in IHDE-I model.
 Also, when we used background datasets alone, the IHDE-II model displayed higher $\xi_{2}$ and $ c_{\rm h}$ values with clustered DE compared to homogeneous DE.
  Lastly, in IHDE-III, the values of $\xi_3$ and $c_{\rm h}$ were sensitive to the choice of dataset and DE assumptions.

Panels of Fig. \ref{fr7} present the $1\sigma$ and $2\sigma$ confidence contours derived from background data for both clustered (blue) and homogeneous (brown) DE scenarios, with corresponding detailed outcomes provided in Table \ref{tabbacnu}. These best-fit values were used to analyze key background parameters including $w_{\rm de}$, $\Delta E$, and $q(z)$, and to compare the models with one another and with $\Lambda$CDM model. These results are depicted in the panels of Fig. \ref{fwd}

The same figure also shows the confidence contours from the combined background and growth rate data for clustered (red) and homogeneous (green) DE scenarios. Table \ref{tabclus} summarizes the resulting best-fit parameters, which we used to evaluate the evolution of the growth rate $f\sigma_{8}(z)$ and its relative deviation, $\Delta f\sigma_{8}(\%)$, from the $\Lambda$CDM prediction. These results are showed in the panels of Fig. \ref{fs8}, showing $f\sigma_{8}(z)$ in comparison with observational data and $\Delta f\sigma_{8}(\%)$ as a function of redshift $z$.

Interpreting the sign of $\Delta f\sigma_{8}(\%)$ allow us to assess how well each model match the growth rate predictions of the $\Lambda$CDM model. A positive (negative) $\Delta f\sigma_{8}(\%)$ indicates a higher (lower) predicted growth rate relative to $\Lambda$CDM. Specifically, the bottom panel of Fig. \ref{fs8} shows that the HDE model closely reproduces the $\Lambda$CDM growth history, while the IHDE-I (IHDE-II) model predicts a faster (slower) growth rate.

Finally, based on the AIC and BIC using background data alone, the ranking of model fit to observations is: HDE, IHDE-III, IHDE-I, and IHDE-II, regardless of homogeneity or DE clustering.
However, when focusing solely on the assumption of DE homogeneity, all models except HDE exhibit a slightly better fit to the observational data than their with clustered DE.
 Additionally, in the IHDE models, those with homogeneous DE demonstrated slightly better alignment with the observational data than the models with clustered DE [see Table \ref{tabbac}].

Furthermore, when we used background data and growth rate jointly, we found that the HDE and IHDE-I models provided a better fit to the observational data than the other models. Specifically, for the clustered DE, the HDE model showed a slightly better fit than the homogeneous DE, and the IHDE-III and IHDE-II models were ordered next in terms of data fit. Thus, the assumptions about DE homogeneity and clustering, along with the datasets used, can influence the ordering of compatibility of models with the observational data [see Table \ref{tabbacgr}].

\appendix
\section{Proof of Eqs. (\ref{ewq2}) and (\ref{ecq3})}\label{appendix.A}
To derive second-order coupled differential equations that describe the evolution of DE and DM based on Eqs. (\ref{eq:stbh}) and (\ref{eq5}), we follow these steps:
first, we manipulate Eq. (\ref{E77}) to obtain the following relation:
\begin{align} \notag
-3\mathcal{H}\frac{\delta p}{\delta \rho}\delta &= -3\mathcal{H}c_{\mathrm{eff}}^{2}\delta -9\frac{\mathcal{H}^{2}}{k^{2}}(1+w_{\mathrm{de}})(c_{\mathrm{eff}}^{2}-c^{2}_{\mathrm{a}})\theta \\
&\simeq -3\mathcal{H}c_{\mathrm{eff}}^{2}\delta \label{Eee}
\end{align}
In the regime of subhorizon scales ($k^2 \gg \mathcal{H}^2$), we can neglect the second term on the right-hand side of Eq. (\ref{Eee}).
 
Secondly, according to Eq.(\ref{E77}), we can express this relation as
\begin{equation}
k^{2}\frac{\delta p}{\delta \rho}\delta = k^{2}c_{\mathrm{eff}}^{2}\delta + 3\mathcal{H}(1+w_{\mathrm{de}})(c_{\mathrm{eff}}^{2}-c^{2}_{\mathrm{a}})\theta \label{Ebb}
\end{equation}
Now, by substituting Eqs.(\ref{Eee}) and ( \ref{Ebb}) into Eqs.(\ref{eq:stbh}) and (\ref{eq5}), we can express them in the following form:
\begin{align}
&\dot\delta +3\mathcal{H}c^{2}_{\mathrm{eff}}\delta -3\mathcal{H}w_{\mathrm{de}}\delta -\frac{Q_{\rm I} }{\bar{\rho}}\delta +(1+w_{\mathrm{de}})\theta =0 ,\label{E12h}\\
&\dot{\theta}+ \Big[\mathcal{H}\left(1-3c^{2}_{\mathrm{eff}}\right)-\frac{Q_{\rm I}}{\bar{\rho}}\Big]\theta  -k^{2}\phi -
\frac{k^{2}\mathrm{c}^{2}_{\mathrm{eff}}}{1+w_{\mathrm{de}}}\delta  =0 . 
\label{eee} 
\end{align}
Moreover, in order to derive Eqs. (\ref{E12h}) and (\ref{eee}) (see also Ref. \citep{Marcondes2016}), we ignore $\delta\bar Q_{\mu}$. 
We remember that Eqs.(\ref{eq:stbh}) and (\ref{eq5}) or their equivalent  Eqs.(\ref{E12h}) and (\ref{eee}) can be used separately for the components of DE and DM. Based on this, we initially utilize Eqs.(\ref{E12h}) and (\ref{eee}) to obtain a second-order equation that describes the evolution of DE perturbations. By eliminating $\theta$ from the system of Eqs.(\ref{E12h}) and (\ref{eee}), we can derive following equation for $\delta_{\mathrm{de}}$ in terms of conformal time:
\begin{align}\label{eww}
& \ddot {\delta}_{\mathrm{de}}+ \mathcal{\tilde A}_{\mathrm{de}}\dot {\delta}_{\mathrm{de}}+ \mathcal{\tilde B}_{\mathrm{de}} {\delta}_{\mathrm{de}}= \mathcal{\tilde S}_{\mathrm{de}}
\end{align}
where the coefficients $ \mathcal{\tilde A}_{\mathrm{de}}$, $ \mathcal{\tilde B}_{\mathrm{de}}$, and $ \mathcal{\tilde S}_{\mathrm{de}}$ are defined as follows:
\begin{eqnarray} \notag
\tilde{\mathcal{ A}}_{\mathrm{de}}&=&\mathcal{H}(1-3w_{\mathrm{de}})+3\mathcal{H}(c^{2}_{\mathrm{a}}-w_{\mathrm{de}})-2\frac{Q_{\rm I}}{{\rho}_{\mathrm{de}}}\\ \notag
 \mathcal{\tilde B}_{\mathrm{de}}&=&3\mathcal{H}^{2}(c^{2}_{\mathrm{eff}}-w_{\mathrm{de}})\Big[1+\frac{\mathcal{\dot H}}{\mathcal{H}^{2}}-3(w_{\mathrm{de}}+c^{2}_{\mathrm{eff}}-c^{2}_{\mathrm{a}})\Big]\\ \notag
&+&k^{2}c^{2}_{\mathrm{eff}}-3\mathcal{H}\dot{w}_{\mathrm{de}}
+\mathcal{H}\Big[3\big (2w_{\mathrm{de}}-c^{2}_{\mathrm{a}}\big)-1\Big]\frac{Q_{\rm I}}{{\rho_{\mathrm{de}}}}\\  \notag 
&+&\Big(\frac{Q_{\rm I}}{\rho_{\mathrm{de}}}\Big)^{2}  
 -\frac{d}{d\eta}\Big(\frac{Q_{\rm I}}{\rho_{\mathrm{de}}}\Big)\\
 \mathcal{\tilde S}_{\mathrm{de}}&=&-(1+w_{\mathrm{de}})k^{2}\phi  \label{err}
\end{eqnarray}
where $- k^{2}\phi $ is expressed by Poisson Eq. (\ref{eq:pois2}).
Likewise, by utilizing Eqs. (\ref{E12h}) and (\ref{eee}), we can derive a second-order equation that describes the evolution of DM perturbations. In this case, we set \(w_{\mathrm{d}} = c^{2}_{\mathrm{eff}} = c^{2}_{\mathrm{a}} = 0\). The resulting equation is obtained as follows:
 \begin{align}\label{eqq}
& \ddot {\delta}_{\mathrm{dm}}+ \mathcal{\tilde A}_{\mathrm{dm}}\dot {\delta}_{\mathrm{dm}}+ \mathcal{\tilde B}_{\mathrm{dm}} {\delta}_{\mathrm{dm}}= \mathcal{\tilde S}_{\mathrm{dm}}
\end{align}
where, in this case, the coefficients $   \mathcal{\tilde A}_{\mathrm{dm}}$, $   \mathcal{\tilde B}_{\mathrm{dm}}$, and $   \mathcal{\tilde S}_{\mathrm{dm}}$ are defined as follows:
\begin{align} \notag
& \mathcal{\tilde A}_{\mathrm{dm}}=\mathcal{H}-2\frac{Q_{\rm I}}{{\rho}_{\mathrm{dm}}}\\ \notag
& \mathcal{\tilde B}_{\mathrm{dm}}=  -\mathcal{H}\frac{Q_{\rm I}}{{\rho_{\mathrm{dm}}}} +\Big(\frac{Q_{\rm I}}{\rho_{\mathrm{dm}}}\Big)^{2}
 -\frac{d}{d\eta}\Big(\frac{Q_{\rm I}}{\rho_{\mathrm{dm}}}\Big)\\ 
& \mathcal{\tilde S}_{\mathrm{dm}}=-k^{2}\phi  \label{euuu}
\end{align}
Additionally, by utilizing the expressions $\frac{d}{d\eta}=a \mathcal{H}\frac{d}{da}$ and $\frac{d^{2}}{d\eta^{2}}=(a\mathcal{H}^{2}+a\mathcal{\dot{H}}) \frac{d}{da} +a^{2}\mathcal{H}^{2}
\frac{d^{2}}{da^{2}}$, along with Eq. (\ref{eq:pois2}), we can represent Eqs. (\ref{eww} ) and ( \ref{eqq}) in terms of the scale factor. Thus, we obtain the following equations:
\begin{align}\label{ewq}
& {\delta}^{ \prime \prime}_{\mathrm{de}}+A_{\mathrm{de}} {\delta}^{ \prime}_{\mathrm{de}}+B_{\mathrm{de}} {\delta}_{\mathrm{de}}=S_{\mathrm{de}}\\
&  {\delta}^{\prime\prime}_{\mathrm{dm}}+{A}_{\mathrm{dm}} {\delta}^{\prime}_{\mathrm{dm}}+ {B}_{\mathrm{dm}} {\delta}_{\mathrm{dm}}= {S}_{\mathrm{dm}} \label{ecq}
\end{align}
where, the prime denotes the derivative with respect to the scale factor. The coefficients $ {A}_{\mathrm{de}} $, $ {B}_{\mathrm{de}} $, and $ {S}_{\mathrm{de}} $ are defined as follows:
\begin{eqnarray} \notag
{A}_{\mathrm{de}}&=&\frac{3}{a}+\frac{H^{\prime}}{H}+\frac{3}{a}(c^{2}_{\mathrm{a}}-2w_{\mathrm{de}})-\frac{2}{a^{2}H}\frac{Q_{\rm I}}{{\rho}_{\mathrm{de}}}\\ \notag
{B}_{\mathrm{de}}&=&\dfrac{3}{a}(c^{2}_{\mathrm{eff}}-w_{\mathrm{de}})\Big[\frac{2}{a}+\frac{ H^{\prime}}{H}-\frac{3}{a}(w_{\mathrm{de}}+c^{2}_{\mathrm{eff}}-c^{2}_{\mathrm{a}})\Big]\\ \notag
&+&\frac{k^{2}c^{2}_{\mathrm{eff}}}{a^{4}H^{2}}-\frac{3}{a}w^{\prime}_{\mathrm{de}}+\frac{1}{a^{3}H^{2}}\Big[3\big (2w_{\mathrm{de}}-c^{2}_{\mathrm{a}}\big)-1\Big]\frac{Q_{\rm I}}{{\rho_{\mathrm{de}}}} \\  
&+&\frac{1}{a^{4}H^{2}}\Big(\frac{Q_{\rm I}}{\rho_{\mathrm{de}}}\Big)^{2}
 -\frac{1}{a^{2}H}\frac{d}{da}\Big(\frac{Q_{\rm I}}{\rho_{\mathrm{de}}}\Big)\\ \notag
{S}_{\mathrm{de}}&=&\frac{3}{2a^{2}} \left(1+w_{\mathrm{de}}\right) \Big[\Omega_{\mathrm{dm}}\delta_{\mathrm{dm}}+
\Omega_{\mathrm{de}}\delta_{\mathrm{de}}\left(1+3c_{\mathrm{eff}}^{2}\right)\Big]\notag   \label{err}
\end{eqnarray}\\
In addition, the coefficients $ {A}_{\mathrm{dm}} $, $ {B}_{\mathrm{dm}} $, and $ {S}_{\mathrm{dm}} $ can be defined as follows:
\begin{align} \notag
&{A}_{\mathrm{dm}}=\frac{3}{a}+\frac{H^{\prime}}{H}-\frac{2}{a^{2}H}\frac{Q_{\rm I}}{{\rho}_{\mathrm{dm}}}\\ \notag
&{B}_{\mathrm{dm}}=-\frac{1}{a^{3}H^{2}}\frac{Q_{\rm I}}{{\rho_{\mathrm{dm}}}}  
+\frac{1}{a^{4}H^{2}}\Big(\frac{Q_{\rm I}}{\rho_{\mathrm{dm}}}\Big)^{2}
 -\frac{1}{a^{2}H}\frac{d}{da}\Big(\frac{Q_{\rm I}}{\rho_{\mathrm{dm}}}\Big)\\ 
&{S}_{\mathrm{dm}}=\frac{3}{2a^{2}}\Big[\Omega_{\mathrm{dm}}\delta_{\mathrm{dm}}+\Omega_{\mathrm{de}}\delta_{\mathrm{de}}\Big]    \label{eroo}
\end{align}
where $ Q_{\rm Ii}, i=1, 2, 3$, $\frac{Q_{\rm Ii}}{\rho_{\mathrm{dm}}}  $, and $  \frac{Q_{\rm Ii}}{\rho_{\mathrm{de}}}$ for models IHDE-I, IHDE-II, and IHDE-III are summarized in Table \ref{tab1222}.
\section*{DATA AVAILABILITY}
The data that support the findings of this study are publicly available. 
\section*{Acknowledgments}
The author sincerely thanks the anonymous referee for insightful comments and suggestions that significantly improved the clarity and quality of this work.

\bibliographystyle{IEEEtran}
\bibliography{wimp}
\end{document}